\documentclass[11pt,tightenlines]{revtex4}
\usepackage{amsfonts}
\usepackage{amsmath}
\usepackage{txfonts}
\usepackage{graphicx}
\usepackage{amssymb}
\usepackage{color}
\usepackage{wrapfig,lipsum,booktabs}
\usepackage{floatrow}

\newcommand{\beginsupplement}{
        \setcounter{table}{0}
        \renewcommand{\thetable}{S\arabic{table}}
        \setcounter{figure}{0}
        \renewcommand{\thefigure}{S\arabic{figure}}
     }
     
\begin{document}
\title{Combinatorics of Place Cell Coactivity and Hippocampal Maps}
\author{Andrey Babichev$^{1,2}$, Daoyun Ji$^{3}$, Facundo M\'emoli$^{4}$ and Yuri Dabaghian$^{1,2*}$}
\affiliation{$^1$Jan and Dan Duncan Neurological Research Institute, Baylor College of Medicine,
Houston, TX 77030, \\
$^2$Department of Computational and Applied Mathematics, 
Rice University, Houston, TX 77005\\
$^3$Department of Neuroscience, Baylor College of Medicine, Houston, TX 77030, \\
$^4$Department of Mathematics, Ohio State University, Columbus, OH 43210\\
E-mail: dabaghia@bcm.edu}

\date{\today}
\begin{abstract}
It is widely accepted that the hippocampal place cells' spiking activity produces a cognitive map of space. 
However, many details of this representation's physiological mechanism remain unknown. For example, it 
is believed that the place cells exhibiting frequent coactivity form functionally interconnected groups---place 
cell assemblies---that drive readout neurons in the downstream networks. However, the sheer number of 
coactive combinations is extremely large, which implies that only a small fraction of them actually gives rise 
to cell assemblies. The physiological processes responsible for selecting the winning combinations are highly 
complex and are usually modeled via detailed synaptic and structural plasticity mechanisms. Here we propose 
an alternative approach that allows modeling the cell assembly network directly, based on a small number of 
phenomenological selection rules. We then demonstrate that the selected population of place cell assemblies 
correctly encodes the topology of the environment in biologically plausible time, and may serve as a schematic 
model of the hippocampal network.
\end{abstract}
\maketitle

\newpage

\section{Introduction}
\label{section:intro}
\begin{figure} 
\includegraphics[scale=0.84]{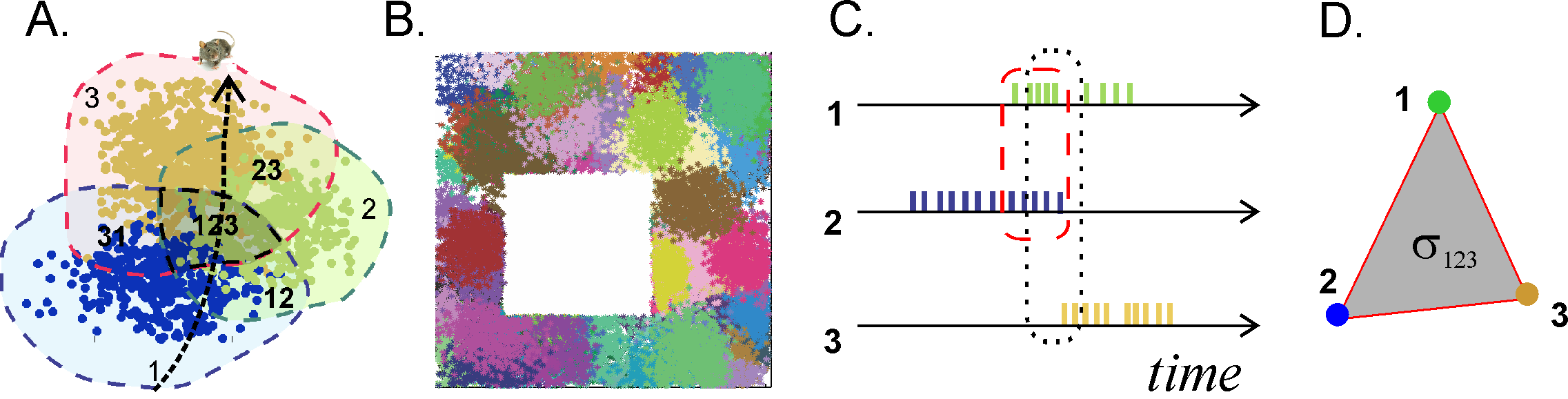}
\caption{\label{PFs} \textbf{Place fields and place cells}. ({\bf A}) The blue, green and brown dots, 
corresponding to the spikes produced by three different place cells, form well-defined spatial clusters, 
which represent their respective place fields. Spikes are positioned in space according to the animal's 
coordinates at the time of spiking. ({\bf B}) A place field map produced by an ensemble of 300 place 
cells with mean peak firing rate $f = 20$ Hz and mean place field size $s = 14$ cm located in a 
$1 \times 1$ m environment. ({\bf C}) A short time segment of the spike trains produced by three place 
cells. The periods of the cells' coactivity, marked by dashed lines, indicate overlap of their respective 
place fields (panel A): cells $c_1$ and $c_2$ are coactive in the region 12, cells $c_1$, $c_2$ and $c_3$ 
are co-active in the region $123$. ({\bf D}) A simplex $\sigma_{123}$ represents schematically the spatial 
connectivity encoded by the coactivity of cells $c_1$, $c_2$ and $c_3$. Its $1D$ edges correspond to 
pairwise coactivity, e.g. $\sigma_{12}$ represents the coactivity of cells $c_1$ and $c_2$.} 
\end{figure}

The mammalian hippocampus plays a major role in spatial learning by encoding a cognitive map of space---a 
key component of animals' spatial memory and spatial awareness \cite{Best1, OKeefe}. A remarkable property 
of the hippocampal neurons---the place cells---is that they become active only in discrete spatial regions---their 
respective place fields \cite{Best2} (Fig.~\ref{PFs}A). A number of studies have demonstrated that place cell 
activity can represent the animal's current location \cite{Zhang,Brown1}, its past navigational experience 
\cite{Carr,Derdikman}, and even its future planned routes \cite{Pfeiffer,Dragoi}. Numerical simulations suggest 
that a population of place cells can also encode a global spatial connectivity map of the entire environment 
\cite{Dabaghian,Arai,Curto}. Hence, it is believed that the large-scale hippocampal representation of space 
emerges from integrating the information provided by the individual place cells, although the details of this 
process remain poorly understood. 

Experimental studies point out that the hippocampal map is topological in nature, i.e., it is more similar to a 
subway map than to a topographical city map \cite{Gothard,Leutgeb,Alvernhe,elife}. In \cite{Dabaghian} we 
proposed a computational approach for modeling its structure based on several remarkable parallels between 
the notions of hippocampal physiology and algebraic topology. For example, \v{C}ech's theorem asserts that the 
topological structure of a space $X$ can be deduced from the pattern of overlaps between regions that cover it 
(for details see \cite{Hatcher} and Methods in \cite{Dabaghian}). The argument is based on building a special 
simplicial complex $\mathcal{N}$, each $n$-dimensional simplex of which corresponds to a nonempty overlap 
of $n+1$ covering regions, and demonstrating that the topological signatures of $\mathcal{N}$ and $X$ are 
same \cite{Hatcher}. Since the place cells' spiking activity induces a covering of the environment by the place 
fields, called a place field map (Fig.~\ref{PFs}B), \v{C}ech's theorem suggests that the place cells' coactivity 
(Fig.~\ref{PFs}C), which marks the overlaps of the place fields, may be used by the brain to represent the 
topology of the environment. In \cite{Dabaghian,Arai,Curto} it was demonstrated that place cell coactivity can 
in fact be used to construct a ``temporal'' analogue of the nerve complex, $\mathcal{T}$, the simplexes of which, 
$\sigma = [c_1, c_2, ..., c_k]$, correspond to the combinations of coactive place cells, $c_1$, $c_2$, ..., $c_k$ 
(Fig.~\ref{PFs}D). Using the methods of persistent homology \cite{Ghrist,Zomorodian} it was shown that the 
topological structure of $\mathcal{T}$ captures the topological properties of the environment, if the range of place 
cell spiking rates and place field sizes happen to parallel biological values derived from animal experiments 
\cite{Dabaghian,Arai}.

However, it remained unclear whether it is possible to physically implement this information in the 
(para)hippocampal network. On the one hand, electrophysiological studies suggest that place cells showing 
repetitive coactivity tend to form so-called cell assemblies---functionally interconnected neuronal groups that 
synaptically drive a readout neuron in the downstream networks \cite{Harris1,Buzsaki,Huyck,Harris2}---which 
may be viewed as ``physiological simplexes'' implementing $\mathcal{T}$. On the other hand, the place cell 
combinations of $\mathcal{T}$ are much too numerous to be implemented physiologically. In a small environment, 
c.a. $1 \times 1$ m, thousands of place cells are active and the activity of 50--300 of them is near maximal level 
at every given location \cite{Buzsaki}. The number of combinations of hundreds of coactive cells in an ensemble 
of thousands is unrealistically large, comparable to $C_{3000}^{100}\sim 10^{200}$. The number of cells in 
most parahippocampal regions, which may potentially serve as readout neurons, is similar to the number of 
place cells \cite{Shepherd}. This implies that only a small fraction of coactive place cell groups may be equipped 
with readout neurons, i.e., that the cell assemblies may encode only a small part of the place cell coactivities---those 
which represent a certain ``critical mass'' of spatial connections.

Physiologically, the place cell assemblies emerge from dynamically changing constellations of synaptic 
connections and are commonly studied in terms of the synaptic and structural plasticity mechanisms 
\cite{Wennekers,Ghalib,Itskov,Caroni,Chklovskii}. For a better qualitative understanding of the qualitative 
properties of the cell assembly network and practical modeling of the hippocampus' functions, we propose 
a biologically plausible empirical approach that allows selecting the most prominent combinations of coactive 
place cells directly and demonstrate that the resulting population of cell assemblies is sufficient for representing 
the topology of the environment.

%%%%%%%%%%%%%%%%%%%%%%%%%%%%%%%%%%%
%%%%%%%%%%%%%%%%%%%%%%%%%%%%%%%%%%%

\section{The Methods}
\label{section:methods}

Mathematically, the task of identifying a subpopulation of coactive place cell combinations corresponds to selecting according 
to biologically motivated criteria a subcomplex $\mathcal{T}_{0}$ of the full \textit{coactivity complex} $\mathcal{T}$. The 
cell assemblies correspond to the \textit{maximal} simplexes of $\mathcal{T}_{0}$, (i.e., the ones that are not subsimplexes 
of any other simplex), in contrast with the maximal simplexes of the coactivity complex, $\mathcal{T}$, which can represent 
any largest combinations of coactive cells. The ``cell assembly complex,'' $\mathcal{T}_{0}$, should satisfy several general 
requirements:

I.	\textbf{Effectiveness}. In the reader–centric approach \cite{Buzsaki}, each cell assembly drives a coincidence detector 
\textit{readout neuron} in the downstream brain regions. Since the number of the readout neurons is comparable to the 
number of place cells, the total number of the maximal simplexes in $\mathcal{T}_{0}$, $N_{\max}(\mathcal{T}_{0})$, 
should be comparable to the number of its vertexes, $N_c(\mathcal{T}_{0})$,
$$N_{\max}(\mathcal{T}_{0}) \approx N_{c}(\mathcal{T}_{0}).$$
However, the algorithm for selecting $\mathcal{T}_{0}$ should reduce only the number of coactive place cell combinations 
and not the place cells themselves, meaning that the number of vertexes in $\mathcal{T}$ and in $\mathcal{T}_{0}$ should 
not differ significantly. In mathematical literature, the number of $k$-dimensional simplexes of a simplicial complex is usually 
denoted as $f_k$, and the list $f$ = ($f_0$, $f_1$, …, $f_d$) is referred to as the complex's $f$-vector \cite{Gromov}. 
However, since in neuroscience literature the letter $f$ is often used to denote firing rates, we denote the number of 
$k$-dimensional simplexes by $N_k$. As a shorthand notation, we use $N_{\max}$ to denote the number of the maximal 
simplexes and $N_c$ the number of $0$-dimensional simplexes in a given complex.

II.	\textbf{Parsimony}. To avoid redundancy, only a few cell assemblies should be active at a given location. Conversely, the 
rat's movements should not go unnoticed by the hippocampal network, i.e., the periods during which all place cell assemblies 
are inactive should be short. 

III.	\textbf{Contiguity}. A transition of the spiking activity from one cell assembly $\sigma_{i}$ to another $\sigma_{i+1}$ 
occurs when some cells in $\sigma_{i}$ shut off and a new group of cells activates in $\sigma_{i+1}$ (see Suppl. Movies). 
The larger is the subassembly $\sigma_{i,i+1} = \sigma_{i}\cap\sigma_{i+1}$ that remains active during this transition (i.e., 
the more cells are shared by $\sigma_{i}$ and $\sigma_{i+1}$) the more contiguous is the representation of the rat's moves 
and hence of the space in which it moves. The overlap between a pair of consecutively active simplexes can be characterized 
by a contiguity index
$$\xi=\frac{\dim(\sigma_i \cap \sigma_{i+1})}{\sqrt{\dim(\sigma_i)\dim(\sigma_{i+1})}},$$
which assumes the maximal value $\xi = 1$ for coinciding cell assemblies and $\xi = 0$ for disjoint ones. In constructing a cell 
assembly complex, we expect that the mean contiguity over the simplexes in $\mathcal{T}_{0}$ should not be lower than in 
$\mathcal{T}$.

\begin{figure}
\floatbox[{\capbeside\thisfloatsetup{capbesideposition={left,top}}}]{figure}[\FBwidth]
%\floatbox[{\capbeside\thisfloatsetup{capbesideposition={left,top},capbesidewidth=4cm}}]{figure}[\FBwidth]
{\caption{\textbf{Topological loops}: each horizontal bar represents the timeline of a topological cycle 
in $\mathcal{T}(T)$: $0D$ loops (connectivity components) and the $1D$ loops. Most cycles last over 
a short time before disappearing. A few remaining, 
\textit{persistent} loops express stable topological information that may correspond to physical obstacles in the rat's environment. 
The time required for the correct number of cycles to appear is interpreted as the minimal time $T_{\min}$ required for the rat 
to learn the environment. The environment used in these simulations (Figure 1B) is topologically connected ($b_0 = 1$), and has 
one central hole ($b_1 = 1$). Thus, the topological barcode of this environment---the list of Betti numbers ($b_0$, $b_1$, $b_2$,...)
---is (1,1, 0, 0, …). The last spurious loop (blue $1D$ loop) disappears at about $T_{\min}= 4.6$ minutes, which is the 
learning time in this case.}\label{Loops}}
{\includegraphics[width=6cm]{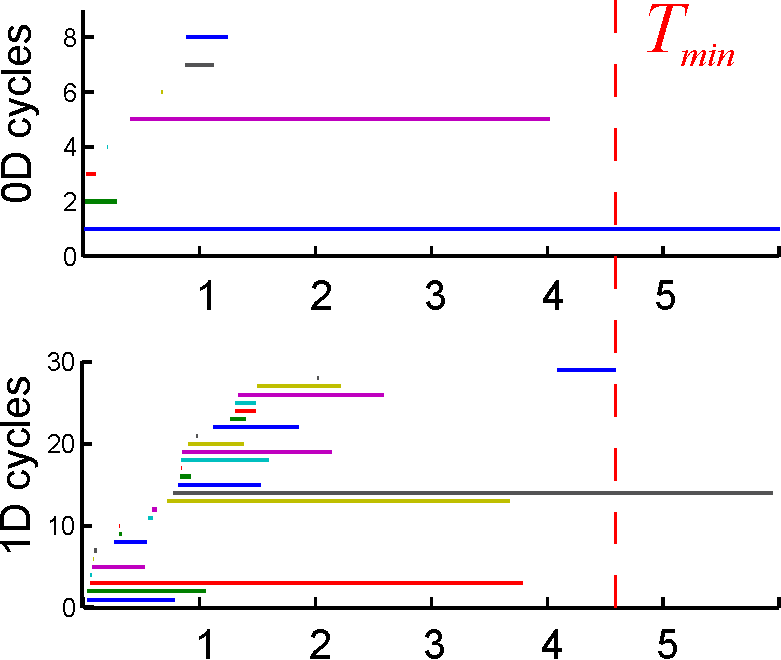}}
\end{figure}

IV.	\textbf{Completeness}. The cell assembly complex $\mathcal{T}_{0}$ should capture the correct topological signatures 
of the environment, such as obstacles, holes, and boundaries. For example, the lowest dimensional $0D$ and $1D$ loops in 
$\mathcal{T}_{0}$ represent, respectively, the piecewise and the path connectivity of the environment, as they are captured 
by the place cell coactivity. This information should emerge from the ``topological noise'' in a biologically plausible time period, 
comparable to the time required to obtain this information via the full complex, $\mathcal{T}$ (see \cite{Dabaghian,Arai} and 
Fig.~\ref{Loops}). 

%%%%%%%%%%%%%%%%%%%%%%%%%%%%%%%%%%%%%%%
%%%%%%%%%%%%%%%%%%%%%%%%%%%%%%%%%%%%%%%

$\vspace{0.5mm}$

\textbf{Place cell spiking} is modeled as a time-dependent Poisson process with spatially localized rate
$$\lambda_c(r)=f_{c}e^{-\frac{(r-r_c)^2}{s^2_{c}}},$$
where $r$ is a point in the environment, $f_c$ is the maximal firing rate of a place cell $c$, and $s_c$ defines the 
size of the corresponding place field centered at $r_c$ \cite{Barbieri}. In a familiar environment, the place fields 
are stable, that is, the parameters $f_c$, $s_c$ and $r_c$ remain constant \cite{Wilson,Brown2}. In our simulations, 
all computations were performed for ten place cell ensembles, each containing 300 neurons with an ensemble mean 
maximal firing rate of 20 Hz and a mean place field size of 30 cm. The place field centers in each ensemble were 
randomly scattered across the environment and most quantities reported in the Results were averaged over ten place 
field configurations.

\begin{figure} 
\includegraphics[scale=0.8]{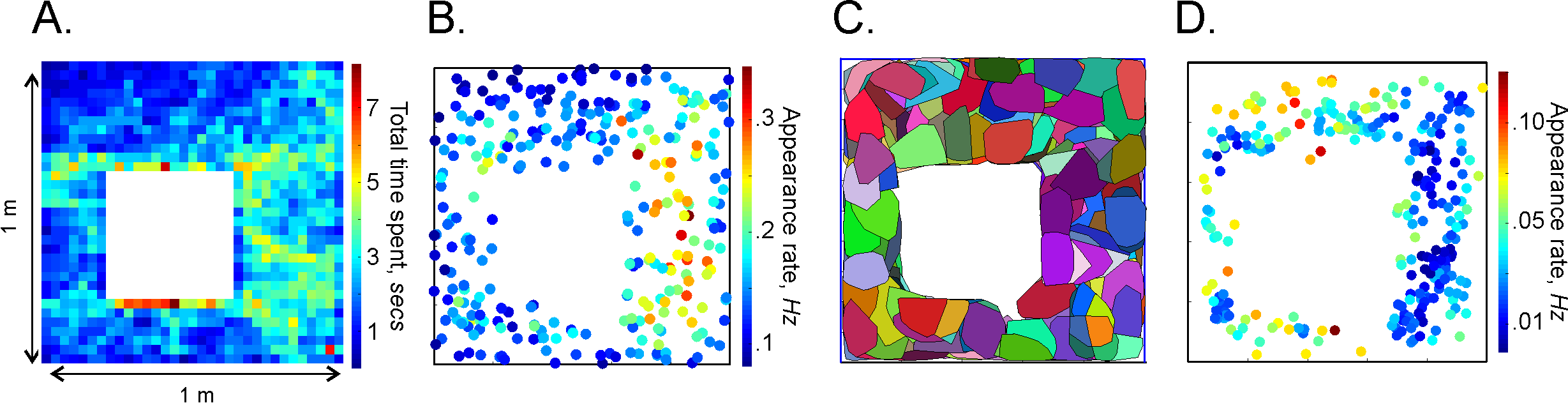}
\caption{\label{Maps} \textbf{Spatial maps}. ({\bf A}) Occupancy of spatial locations in a $1 \times 1$ m environment---a 
$2D$ histogram of the time spent by the animal in different locations. ({\bf B}) Frequency of place cells' spiking: each dot 
marks the location of a place cell's center $r_c$ and indicates the corresponding appearance rate according to the colorbar. 
Higher appearance rates appear in the domain where the spatial occupancy is higher. ({\bf C}) Simplex field map. The place 
field map for the same place cell ensemble is shown in Fig.~\ref{PFs}B. ({\bf D}) Spatial distribution of the frequency of the 
maximal simplexes' appearances. Notice that, since place cells with higher appearance rates tend to produce higher order 
cell assemblies, which, in turn, have lower appearance rates, the spatial distribution of rates on B. and D. are complementary.} 
\end{figure}

\textbf{Spatial map}. We simulated the rat's movements through a small ($1\times 1$ m) planar environment (Fig.~\ref{PFs}1), 
similar to the arenas used in typical electrophysiological experiments (see Methods in \cite{ Dabaghian}) over $T = 25$ 
minutes---the duration of a typical ``running session.'' The spatial occupancy rate of the rat's trajectory (i.e., the histogram 
of times spent at a particular location) and the frequency of the place cells' activity are shown on Fig.~\ref{Maps}A,B. 
The mean speed of the rat is 20 cm/sec, so that turning around the central obstacle takes about 7 seconds.

%%%%%%%%%%%%%%%%%%%%%%%%%%%%%%%%%%%
%%%%%%%%%%%%%%%%%%%%%%%%%%%%%%%%%%%

By analogy with the place fields, we designate the spatial domain where a combination of place cells comprising a 
simplex $\sigma$ is active as its \textit{simplex field}, $s_{\sigma}$ (Fig.~\ref{Maps}C). If the simplex corresponds 
to a cell assembly, then $s_{\sigma}$ may also be referred to as the \textit{cell assembly field}. Similarly to the place 
fields and the place field map (Fig.\ref{PFs}B), the collection of all simplex fields forms a \textit{simplex field map} and 
the cell assembly fields form a \textit{cell assembly map} ((Fig.~\ref{Maps}C). These maps provide a better ``geometric 
proxy'' for the rat's cognitive map because they illustrate both the activity and the \textit{co}activity of the individual place 
cells ((Fig.~\ref{Maps}C-D). In the following, the structure of these maps will be used to discuss our selection algorithms. 
If the distinction between a cell assembly map and a simplex map is not essential, it will be referred to as a space map.

\begin{figure}
\floatbox[{\capbeside\thisfloatsetup{capbesideposition={right,top}}}]{figure}[\FBwidth]
%\floatbox[{\capbeside\thisfloatsetup{capbesideposition={left,top},capbesidewidth=4cm}}]{figure}[\FBwidth]
{\caption{\textbf{An activity raster} of a population of 20 place cells over 250 time bins. Each row defines 
the activity vector of the corresponding place cell. The color of the ticks indicates the number of spikes contained in the 
corresponding bin, according to the colorbar on the right. At every time step, the nonempty bins in the vertical column 
define the list of currently active cells, i.e., the active simplex $\sigma_t$. During the time interval $T$, marked by the 
two vertical blue lines, cell $c_{16}$ is coactive with $c_{18}$ but not coactive with $c_{14}$.}\label{Raster}}
{\includegraphics[scale=0.7]{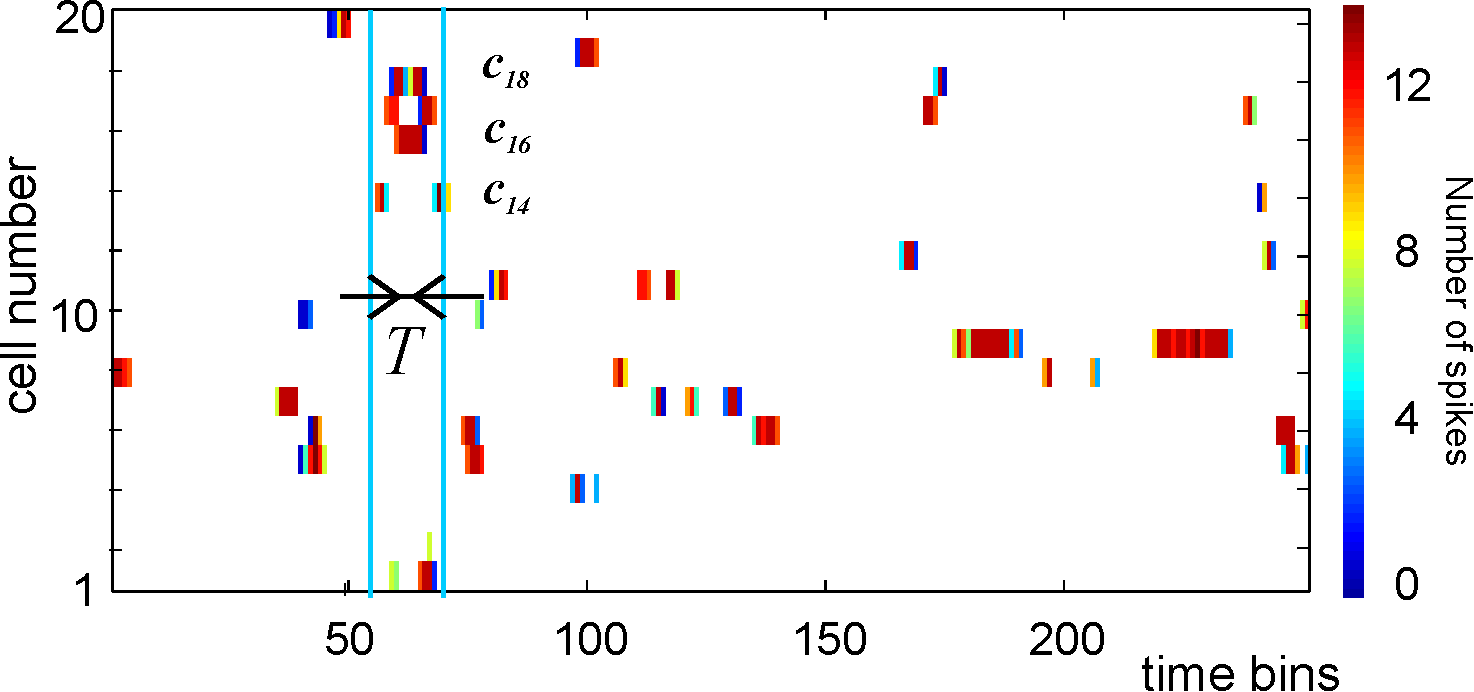}}
%{\includegraphics[width=6cm]{Figure4}}
\end{figure}

\textbf{Population activity}. To define the population code \cite{Pouget} of place cell combinations, we construct place cell 
\textit{activity vectors} by binning spike trains into 1/4 sec long time bins (for a physiological justification of this value see 
\cite{Arai,Mizuseki}). If the time interval $T$ splits into $n$ such bins, then the activity vector of a cell $c$ is
$$m_{c}(T) = [m_{c,1}, …, m_{c,n}],$$
where $m_{c,k}$ specifies how many spikes were fired by $c$ in the $k^{th}$ time bin. The components of $m_{c}$, 
normalized by the total number of spikes, $M_{c}$, define spiking probabilities, $p_{c,k} = m_{k}/M_{c}$ \cite{Perkel}. 
A stack of activity vectors forms an \textit{activity raster} illustrated on Fig.~\ref{Raster}.

Two cells, $c_1$ and $c_2$, are \textit{coactive} over a certain time period $T$, if the dot product of their activity vectors 
does not vanish,
$$m_{c_1}(T) \cdot m_{c_2} (T) \neq 0.$$
The component-wise or Hadamard product of two activity vectors
$$m_{c_1, c_2}=m_{c_1}\odot m_{c_2}=[m_{c_1,1}m_{c_2,1}, m_{c_1,2}m_{c_2,2},…,m_{c_1,n} m_{c_2,n}]$$
defines the \textit{coactivity vector} of cells $c_1$ and $c_2$, which can also be viewed as the activity vector of the 
corresponding $1D$ simplex $\sigma_{12} = [c_1, c_2 ]$, $m_{\sigma_{12}}\equiv m_{c_1,c_2}$. Similarly, the 
Hadamard product of $k$ vectors, 
$$m_{\sigma_{12...k}}=m_{c_1 , c_2 , ...,c_k} = m_{c_1} \odot m_{c_2} \odot … \odot m_{c_k},$$
defines the \textit{activity vector of the simplex} $\sigma_{12...k} = [c_1, c_2, …, c_k]$.

For each activity vector, $m_{\sigma}$, we also define its bit array mapping into a binary \textit{appearance vector}, 
$a_{\sigma}$, which indicates during which time-bins the corresponding simplex $\sigma$ has made its appearance, 
i.e., $a_{\sigma,i} = 1$ iff $m_{\sigma,i} > 0$. The \textit{appearance rate}, $f_{\sigma}(T)$, of a simplex $\sigma$ 
over a time interval $T$, is defined as the $L_{1}$ norm of its appearance vector, averaged over that time interval,
$$f_{\sigma}(T) = (1/T)\Sigma_{i} a_{\sigma,i}.$$
These appearance vectors and appearance rates allow distinguishing the intrinsic physiological characteristics of place 
cells' spiking, e.g., their maximal firing rate, from the frequency with which these cells activate due to the rat's movements 
through their respective place fields. While the maximal firing rate of a typical place cell is about 15 Hz \cite{Best1}, the 
frequency of their activation is much lower.
%%%%%%%%%%%%%%%%%%%%%%%%%%%%%%%%%%%%%%%
%%%%%%%%%%%%%%%%%%%%%%%%%%%%%%%%%%%%%%%

\section{Results}
\label{section:results}

The simulated ensembles of 300 place cells in the environment shown on Fig.~\ref{PFs}B produced a coactivity complex 
$\mathcal{T}$ with about $N_{\max} = 1000$ maximal simplexes. Despite the high dimensionality of these simplexes 
(up to $D = 35$, mean $\bar D = 17$), the characteristic dimensionality of a facet shared by two consecutively active 
simplexes, $\sigma_{i}$ and $\sigma_{i+1}$, is relatively low, so that the mean contiguity of $\mathcal{T}$ is $\xi = 0.6$. 
This implies that, geometrically, if the simplexes of $\mathcal{T}$ are viewed as multidimensional tetrahedrons, the 
selected complex, $\mathcal{T}_{0}(\theta)$, assumes a highly irregular shape (Fig.~\ref{SupplFigure1}A).

\begin{figure} 
\includegraphics[scale=0.8]{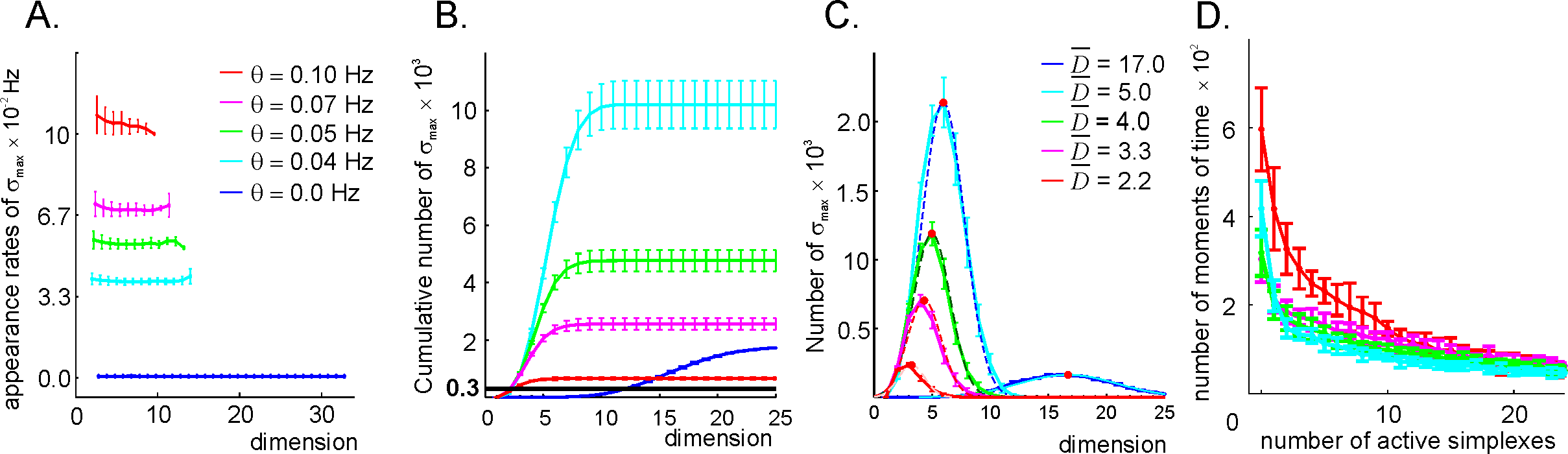}
\caption{\label{M0} \textbf{A direct selection of the simplexes by appearance rates}. ({\bf A}) In the original coactivity 
complex $\mathcal{T}(\theta = 0)$, the maximal simplexes $\sigma_{\max}$ appear on average but once during the 
entire observation period, resulting in low appearance rates ($f_{\sigma} <10^{-3}$ Hz, blue line). Imposing four 
different thresholds $\theta$ (color coded) raises the appearance rates of the selected maximal simplexes almost 
uniformly in all dimensionalities. ({\bf B}) Cumulative distribution of the number of maximal simplexes $N_{\max}$ 
over the selected simplexes' dimension. In the $\mathcal{T}(\theta = 0)$ case $N_{\max}$  exceeds  the number 
of  vertexes ($N_c = 300$, black horizontal line) by almost an order of magnitude. Small threshold values result in 
an explosive increase of $N_{\max}$ which then begins to decrease for $\theta > 0.04$ Hz, remaining significantly 
higher than $N_c$ for all four tested values of $\theta$. ({\bf C}) The histograms of the maximal simplexes' 
dimensionalities fit with normal distribution. The high mean dimensionality ($\bar D = 17$) observed in the 
$\mathcal{T}(\theta = 0)$ case reduces to $\bar D = 2.2$ for $\theta = 0.1$ Hz. The width of the distributions is 
about $50\% $ of $\bar D$. ({\bf D}) The histograms of the number of the coactive maximal simplexes, fit to an 
exponential distribution demonstrate that the typical number of coactive simplexes is large, $\beta >10$. All values 
are averaged over ten place field maps generated by ten place cell ensembles with the same mean peak firing rate 
and mean place field size.} 
\end{figure}

More importantly, nearly $100 \% $ of the maximal simplexes appeared only once during the entire 25 minute period 
of navigation, i.e., a typical maximal simplex's appearance rate is low, $f_{\sigma} \sim 10^{-3}$ Hz. However, a 
typical vertex activated about 200 times or every seven seconds, suggesting that some of the lower dimensional 
subsimplexes may be better candidates for forming cell assemblies. Is it then possible to build a cell assembly 
complex $\mathcal{T}_{0}$ by discarding the high-dimensional maximal simplexes with low appearance rates and 
retaining their subsimplexes that appear more frequently? We tested this hypothesis by identifying the combinations 
$\sigma$ whose coactivity exceeds a certain threshold $f_{\sigma} > \theta$, and studied the properties of the 
resulting simplicial complex as a function of $\theta$ (Fig.~\ref{M0}A).

%%%%%%%%%%%%%%%%%%%%%%%%%%%%%%%%%%%
%%%%%%%%%%%%%%%%%%%%%%%%%%%%%%%%%%%

First we observed that, as soon as the appearance threshold is introduced ($\theta \gtrsim 10^{-3}$ Hz), the high 
dimensional simplexes start braking up, releasing large numbers of lower dimensional subsimplexes: the number of 
$k$-dimensional subsimplexes in a $n$-dimensional simplex grows as combinatorial coefficient $C_{n+1}^{k+1}$, 
e.g., for $n = 17$ and $k = 7$, $C_{18}^{8} \approx 44,000$. As a result, the complex $\mathcal{T}_{0}(\theta)$ 
rapidly inflates. As $\theta$ increases further ($\theta > 0.04$), the number of ``passing'' simplexes decreases, and 
$\mathcal{T}_{0}(\theta)$ begins to shrink in all dimensions (i.e., $N_{D}(\theta_1) > N_{D}(\theta_2)$ for 
$\theta_1 < \theta_2$, for all $D$, (Fig.~\ref{M0}B). Despite this, the number of maximal simplexes remains 
high: $N_{\max} = 30 \times N_{c}$ at $\theta = 0.04$ Hz, $N_{\max} = 7 \times N_{c}$ at $\theta = 0.07$ Hz and 
$N_{\max} = 3 \times N_{c}$ for the highest tested threshold, $\theta = 0.1$ Hz (Fig.~\ref{M0}B), while their 
characteristic dimensionality drops from $D = 17$ to $D = 7$ at $\theta = 0.04$ Hz and to $D = 3$ at $\theta = 0.1$ Hz. 
The mean contiguity index for this range of thresholds remains close to $\xi = 0.7$, indicating that the degree of 
overlap between the selected combinations of place cells is higher than in the original coactivity complex.

However, raising the passing threshold $\theta$ quickly destroys the geometric integrity of the resulting complex's 
spatial map. As shown on Fig.~\ref{SupplFigure2}, for $\theta = 0.05$ Hz, only $\sim 50\% $ of the environment 
is covered by the remaining simplex fields, and for $\theta = 0.07$ Hz the simplex map barely retains its one-piece 
connectedness: in some cases the complex $\mathcal{T}_{0}$ splits in two (the corresponding Betti numbers, $b_0$, 
are listed in Table~\ref{SupplTable1}, for an illustration see Fig.~\ref{SupplFigure3}). For $\theta = 0.1$ Hz, the complex 
fragments into multiple components (mean $b_0 \sim 7$) that are riddled with holes: the Betti numbers $b_{n>0}$ 
indicate the presence of hundreds of stable loops in higher dimensions. Thus, even if the coactive place cell combinations 
selected at $\theta \geq 0.05$ Hz could be supplied with readout neurons and would form cell assemblies, the resulting 
cell assembly network would not encode the correct spatial connectivity.

An additional problem is that reducing the order of the assemblies violates the ``assembly code'' for spatial locations: 
every time several subsimplexes $\sigma_{i}$ are selected from a high-order maximal simplex $\sigma$, several 
overlapping simplex fields $s_{\sigma_i}$ are produced in place of a single $s_{\sigma}$. As a result, the parsimony 
of the representation is compromised: a location that was previously represented by a single simplex becomes 
represented by a few of its subsimplexes (Fig.~\ref{SupplFigure4}A-B). Fig.~\ref{M0}D shows a histogram of 
the numbers of simultaneously active maximal simplexes in $\mathcal{T}_{0}$: although most of the time only a 
few maximal simplexes are active, a coactivity of many of them ($n > 25$) is not uncommon. Conversely, while most 
of the time---on average $84\% $ for the selected place cell ensembles---at least one simplex is active, longer inactivity 
periods are observed as described by double exponential distributed with the rate $\beta \approx 3.5$ sec (Fig.~\ref{SupplFigure2}).

Overall, since most of the $\mathcal{T}_{0}$-requirements listed in the Methods fail, we are led to conclude that the 
most straightforward selection rule, based on selecting high appearance rates, does not produce the desired trade–off 
between the order of the assemblies, the frequency of their appearances, and the quality of topological representation 
of the environment. This failure motivates the search for alternative methods.

\textbf{Method I}. To produce a more detailed approach to selecting coactive cell combinations, we observe that place 
fields are typically convex planar regions, and hence the existence of higher order overlaps between them actually follows 
from the lower order overlaps. According to Helly's theorem, a collection of $n > D +1$ convex $D$-dimensional regions 
in Euclidean space $R^{D}$ will necessarily have a nonempty common intersection, if the intersection of every set of 
$D + 1$ regions is nonempty (see \cite{Avis,Eckhoff} and Fig.~\ref{SupplFigure4}D). From the perspective of \v{C}ech 
theory, this implies that if $n$ convex regions which cover a $D$-dimensional space contribute all the combinatorially possible 
$D$-dimensional simplexes to the nerve complex, then they also provide all the higher (up to $n-1$) dimensional simplexes 
to it. In a planar ($D = 2$) environment, this implies that a set of four or more place fields has a common intersection, 
if any three of them overlap. Moreover, although mathematically it is possible that three place fields exhibit pairwise, but 
not triple overlap, the probability of such an occurrence is low (Fig.~\ref{SupplFigure4}C). A direct computational 
verification shows that if a triple of place cells demonstrates pairwise coactivity, then, in over $90\% $ of cases, it also 
correctly encodes a triple spatial overlap. In other words, a ``clique'' of pairwise coactivities indicates the overlaps of all 
higher orders, which implies that the spatial connectivity graph $G_{\mathcal{N}}$ whose vertexes correspond to the place 
fields and links represent pairwise overlaps, encodes most simplexes in the nerve complex $\mathcal{N}$. 

As a reminder, a clique of an undirected graph is a set of pairwise connected vertices. From the combinatorial perspective, 
a clique and a simplex have the same defining property: any subset of a simplex is its subsimplex and any subset of a 
clique is its subclique; a maximal clique is the one that is contained in no other clique. Hence, each graph defines its own 
``clique complex,'' the $k$-dimensional simplexes of which corresponds to the graph's cliques with $k+1$ vertices \cite{Bandelt}.

The observation that the nerve complex induced from the place field map can be approximated by the clique complex of 
the place field pairwise connectivity graph, suggests that the corresponding coactivity complex $\mathcal{T}$ can also be 
built based only on pairwise, rather than higher-order, coactivities. This approach is well justified physiologically, since 
pairwise coactivity detector pairs of synapses are commonly observed \cite{Katz,Brette}. The rule for defining the temporal 
analogue of $G_{\mathcal{N}}$---the \textit{relational graph} $G_{\mathcal{T}}$---is straightforward: a pair of vertexes is 
connected in $G_{\mathcal{T}}$ if the corresponding cells $c_i$ and $c_j$ are coactive. Thresholding pairwise coactivity 
rates according to the rule
\begin{equation}
C_{ij} = \begin{cases} 1 &\mbox{if } f_{c_i,c_j} \geq \theta \\ 
0 & \mbox{if }  f_{c_i,c_j} <\theta. 
\end{cases}
\label{C}
\end{equation}
allows constructing a family of relational graphs $G_{\mathcal{T}(\theta)}$ over the pairs of place cells with high coactivity. 
The higher the threshold is, the sparser its connectivity matrix $C_{ij}$ and the smaller the number of maximal cliques and 
hence of maximal simplexes in the corresponding clique complex. Since in the following the graph $G_{\mathcal{N}}$ will 
not be used we will suppress the subscript ``$\mathcal{T}$'' in the notation for $G_{\mathcal{T}}$.

%%%%%%%%%%%%%%%%%%%%%%%%%%%%%%%%%%%
\begin{figure} 
\includegraphics[scale=0.8]{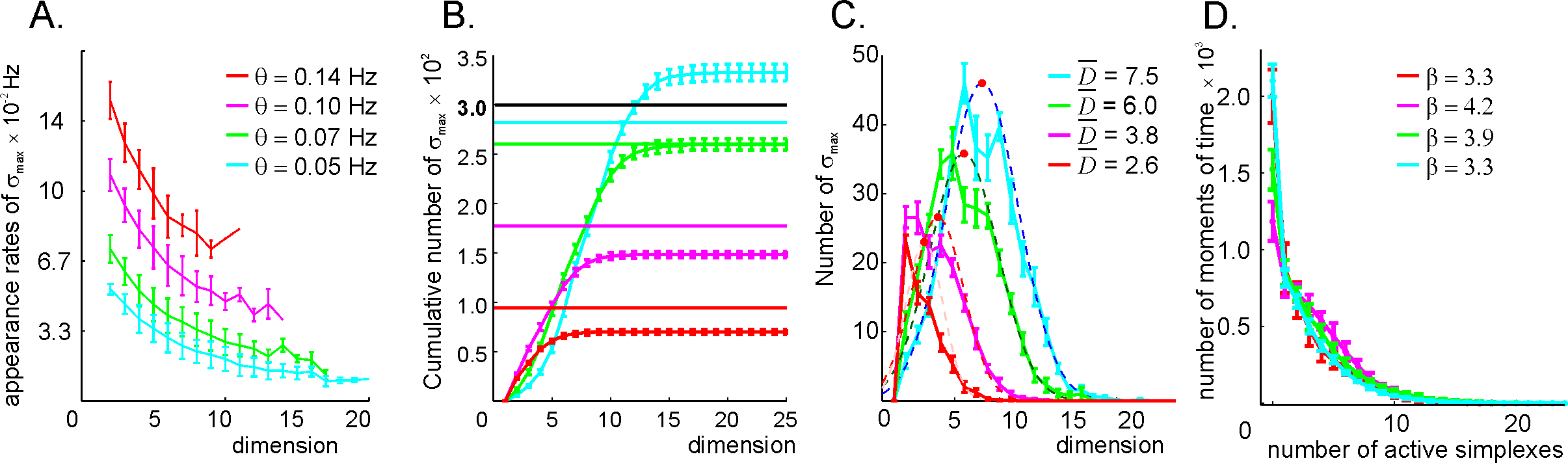}
\caption{\label{M1} \textbf{Selecting maximal simplexes via the pairwise coactivity threshold (Method I)}. ({\bf A}) The 
appearance rates of the maximal simplexes computed for four different pairwise appearance rate thresholds $\theta$ 
decrease as a function of their dimensionality. The values at $D = 1$ correspond to the value of the threshold imposed 
on the links' appearance rate. ({\bf B}) Cumulative distribution of the numbers of maximal simplexes, $N_{\max}$, over 
the selected simplexes' dimension. The numbers of cells $N_c$ for each threshold value are shown by horizontal lines. 
The tendency of the maximal simplexes to outnumber the vertexes $N_{\max} > N_c$, characteristic for small values of 
$\theta$, is reversed around $\theta = 0.07$ Hz, where $N_{\max}$ and $N_c$ level out. ({\bf C}) The histograms of 
the maximal simplexes' dimensionalities fit with normal distribution. The mean dimensionalities are similar to the ones 
produced by the previous selection method. The width of the distributions is about $50 \% $ of $\bar D$. ({\bf D}) The 
histogram of the number of coactive maximal simplexes, fit to an exponential distribution, shows that the expected 
number of coactive simplexes ($\beta \sim 4$) is significantly lower than in the previous selection method. The procedure 
of averaging over the place field maps is the same.} 
\end{figure} 
%%%%%%%%%%%%%%%%%%%%%%%%%%%%%%%%%%%
We studied the relational graphs $G(\theta)$ and their respective clique complexes 
$\mathcal{T}_{0}(G(\theta))\equiv \mathcal{T}_{0}(\theta)$ as a 
function of $\theta$. First, we observed that the appearance rates of the maximal simplexes in $\mathcal{T}_{0}(\theta)$ 
become sensitive to the simplexes' dimensionality (Fig.~\ref{M1}A), implying that this selection procedure in effect 
attributes different thresholds to simplexes of different dimensions by using only one free parameter $\theta$. Second, 
the size of $\mathcal{T}_{0}(\theta)$ is not as large as before. As shown on Fig.~\ref{M1}B, even for a relatively 
low threshold $\theta = 0.05$ Hz, the number of maximal simplexes exceeds the number of cells only marginally. For 
higher thresholds, this number steadily decreases: $N_{D}(\theta_1) < N_{D}(\theta_2)$ for $\theta_1 > \theta_2 > 0.04$ 
and $D > 3$, though in lower dimensions ($1\leq D \leq 3$) this number may increase. The characteristic contiguity ranges 
between $\xi = 0.65$ at $\theta = 0.05$ Hz to $\xi = 0.72$ at $\theta = 0.14$ Hz, which is higher than the value produced 
by the direct simplex selection method. Geometrically, this implies that the collection of maximal simplexes selected by 
pairwise threshold selection is more aggregated than the collection produced via direct simplex selection, i.e., the resulting 
complex $\mathcal{T}_{0}(\theta)$ is geometrically more similar to a ``simplicial quasimanifold'' (see Fig.~\ref{SupplFigure1}B). 
However, the number of place cells $N_{c}$ drops as a result of discarding too many links with low appearance rate: 
$N_c = 290$ at $\theta = 0.05$ Hz and $N_c = 100$ at $\theta = 0.14$ Hz. At $\theta = 0.1$ Hz number of cells levels 
out with the number of maximal simplexes, $N_{\max} \sim N_c = 260$.

As before, raising the coactivity threshold degrades the spatial map. At $\theta > 0.07$ Hz the simplex fields 
no longer cover the environment and at $\theta > 0.1$ Hz the map fragments into pieces (Fig.~\ref{SupplFigure5}). 
However the resulting complex exhibits a much more regular topological behavior: the correct signature ($b_0 = 1$, 
$b_1 = 1$, $b_2 = 0$, $b_3 = 0$, ...) in $\mathcal{T}_{0}(\theta)$ appears at $\theta = 0.05$ Hz. The higher order 
Betti numbers ($b_{n \geq 2}$) remain trivial at still higher $\theta$s (Table~\ref{SupplTable2}A), 
even though the connectedness and path connectivity of the environment ($b_1$ and $b_0$) become misrepresented.

%%%%%%%%%%%%%%%%%%%%%%%%%%%%%%%%%%%
\begin{figure} 
\includegraphics[scale=0.84]{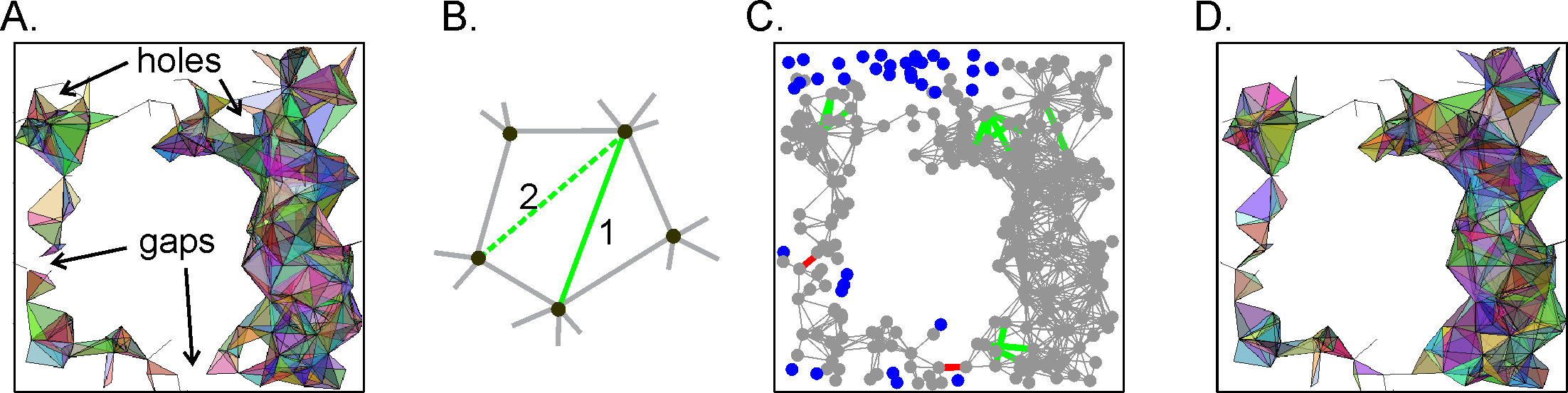}
\caption{\label{Corrections} \textbf{Figure 7. Correction algorithms}. ({\bf A}) A spatial projection of the $2D$ skeleton 
of $\mathcal{T}(\theta)$ shows gaps and holes that compromise, respectively, the piecewise and path connectivity of 
$\mathcal{T}(\theta)$. If the links across the gaps and holes of $\mathcal{T}(\theta)$ are restored, then its correct 
connectivity structure may be regained. ({\bf B}) A `hole' produced by five connected vertexes is closed by restoring 
some of the previously discarded crosslinks. ({\bf C}) A projection of the relational graph $G$ into the environment, 
shown in grey. The edges added across the gaps are shown in red and the edges added to fill the holes are shown in 
green. The vertexes that are left disconnected due to low appearance rates of the edges connecting them to other 
vertexes are shown by blue dots. ({\bf D}) A spatial map of the resulting ``patched'' $2D$ skeleton of $\mathcal{T}(\theta)$. 
The parameter values are $n_g =15$, $m_h = 10$, and the lowered threshold for reintroducing the missing links is $\theta = 50$.} 
\end{figure} 
%%%%%%%%%%%%%%%%%%%%%%%%%%%%%%%%%%%

This improvement of the behavior of $\mathcal{T}_{0}(\theta)$ suggests that, despite all the shortcomings, 
the link-selection strategy may lead to a successful model of the place cell assembly network. After all, it is not 
surprising that a single selection rule does not resolve all the aspects of the cell assembly formation.  Yet if it 
captures the essence of the process, it should be possible to correct or to adjust its outcome. For example, 
one of the difficulties faced by the coactivity selection algorithm is that, for high $\theta$, $\mathcal{T}_{0}(\theta)$ 
may brake into several pieces. However, the gaps between them are small. Thus, if a few discarded edges of the 
relational graph that originally bridged these gaps are retained, then the connectedness of $\mathcal{T}_{0}(\theta)$ 
may be spared (Fig.~\ref{Corrections}A). Similarly, a ``hole'' in the relational graph is a linear chain of edges, 
connected tail to tail, with no shortcuts. However, if the links with the lower appearance rate 
($f\geq \theta_h$, $\theta_h < \theta$) that span across the hole exist at $\theta = 0$, then they also can 
be restored (Fig.~\ref{Corrections}B). This may remove the non-contractible chains of $1D$ simplexes in 
$\mathcal{T}_{0}(\theta)$ that compromised its path connectivity (Fig.~\ref{Corrections}C,D). Thus, we implemented 
the following two \textit{rectification algorithms}:
\begin{enumerate}
\item \textbf{Filling gaps}: find pairs of vertexes $v_a$ and $v_b$ separated in $G(\theta)$ by more than $n_g$ 
edges and then test whether these vertexes are connected directly by links (from $G(\theta = 0))$ whose 
appearance rate exceeds a lower threshold $\theta_g < \theta$. If such links exist, add them to $G(\theta)$ (red 
lines on Fig.~\ref{Corrections}C).
\item \textbf{Closing holes}: A closed chain containing $m_h \geq 4$ edges in $G(\theta)$, with no shortcuts, is likely 
to produce a hole in $\mathcal{T}_{0}(\theta)$. We identified such chains and restored the discarded cross-links whose 
appearance rate exceeds a lower threshold $\theta_h < \theta$.
\end{enumerate}

Thus, both rectification algorithms depend on two parameters: the length of the involved chains ($n_g$ for gaps and 
$m_h$ for holes) and the value of the reduced threshold $\theta_g$ and $\theta_h$. In our numerical experiments, we 
found that the optimal value for the thresholds is $\theta_h = \theta_g = 50$, and the parameters range between 5 
and 10 ($m_h$) and 10 and 15 ($n_g$). Typically, each rectification procedure is applied once or twice before the 
right signature of $\mathcal{T}_{0}(\theta)$ is achieved, and this without producing significant changes of the complex's 
structure, such as altering the appearance of its simplexes or increasing its size $N_{\max}$. As illustrated in the 
Table~\ref{SupplTable2}B, the correct signature in the ``repaired'' complex is achieved for all cases at $\theta = 0.07$ Hz. 
In particular, at $\theta = 0.07$ Hz we obtain a simplicial complex $\mathcal{T}_{0}$ with the correct signature, having 
$N_c = 260$ vertexes and about the same number of maximal simplexes, $N_{\max} \approx N_c$. These maximal 
simplexes appear on average at a rate of $f_{\sigma} \geq 0.07$ Hz, at least during every other run of the rat around 
the environment, and have dimensionality $D = 6$. As a result, the requirements to $\mathcal{T}_{0}$ are met and the 
maximal simplexes of $\mathcal{T}_{0}$ may represent hippocampal place cell assemblies that together encode a map 
of the environment, and hence $\mathcal{T}_{0}$ itself can be viewed as the ``cell assembly complex.''

%%%%%%%%%%%%%%%%%%%%%%%%%%%%%%%%%%%
\begin{figure} 
\includegraphics[scale=0.77]{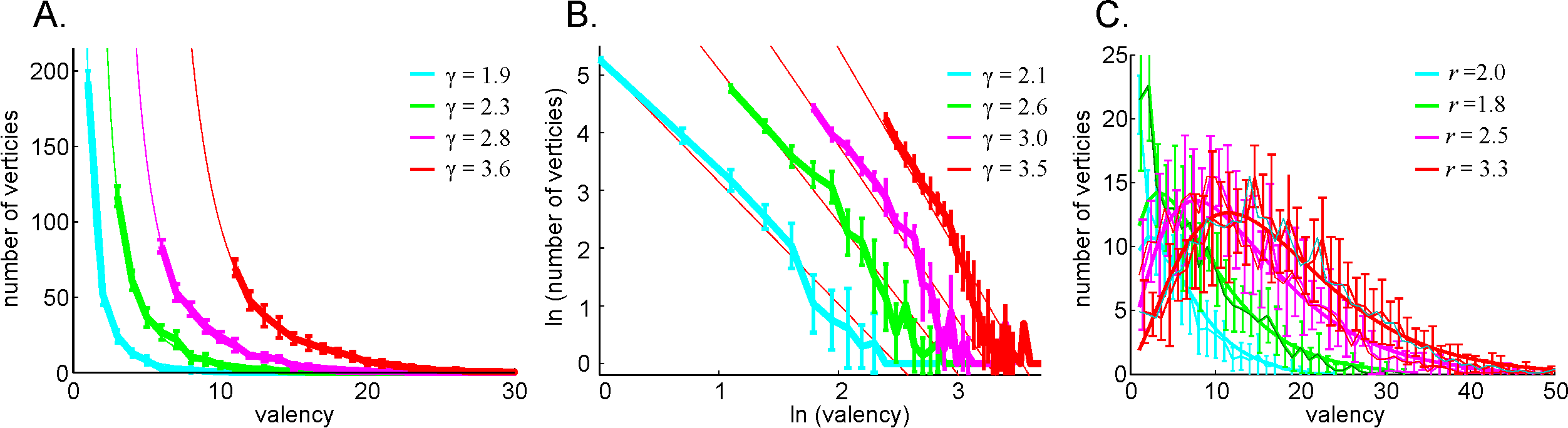}
\caption{\label{Statistics} \textbf{Statistics of the vertex degrees in relaitonal graphs}. ({\bf A}) The histogram 
of the vertex degrees $k$ in the neighbor-controlled relational graph $G(n_0)$, computed for four different $ n_0$ (Method II) 
and fitted to a power law distribution $P(k) \sim k^{-\gamma}$. The graph demonstrates that $G(n_0)$ is a scale-free network. 
({\bf B}) The same distribution on the log-log scale and an independent linear fit of the powers $\gamma$. The confidence 
intervals of the two fits, ranging between $\pm 0.15$ and $\pm 0.3$, overlap for each case. ({\bf C}) In the pairwise coactivity 
threshold (Method I), the histogram of the relational graph's vertex degrees is fit by negative binomial distribution, suggesting 
that $G(\theta)$ is similar to a random network.} 
\end{figure} 
%%%%%%%%%%%%%%%%%%%%%%%%%%%%%%%%%%%

\textbf{Method II}. A common feature of the appearance-rate-based selection rules is that the resulting simplicial complex 
reflects biases of its spatial occupancy: higher dimensional maximal simplexes concentrate over the parts of the environment 
where the rat appears more frequently. In particular, the relational graph shows a higher concentration of edges over the 
eastern segment of the environment (Fig.~\ref{Corrections}C) where the occupancy rate is highest (Fig.~\ref{Maps}A). 
On the one hand, this is natural since the frequency of the place cells' spiking activity certainly does depend on the frequency 
of the rat's visits to their respective place fields, which therefore affects the hippocampal network's architecture 
\cite{Caroni,Chklovskii}. In fact, this argument is at the core of the classical ``hippocampus as a cognitive graph'' model 
\cite{Muller,Burgess}, which proposes that the architecture of the hippocampal network is an epiphenomenon of the place 
cell coactivity. On the other hand, the physiological processes that produce synaptic connections may be more autonomous. 
For example, the CA3 region of the hippocampus is anatomically a recurrent network of place cells whose spiking activity 
and synaptic architecture are dominated by the network's attractor dynamics \cite{Colgin,Wills,Tsodyks}. 

These considerations lead us to test an alternative method of constructing the relational graph based on selecting, for every 
cell, its $n_0$ closest neighbors as defined by the pairwise coactivity rate $f_{c_i,c_j}$. Note that the resulting number of 
connections may be different for different cells: a cell $c_1$ may be among the $n_0$ closest neighbors of a cell $c_2$, and 
hence $c_1$ and $c_2$ become connected, but the set of $n_0$ closest neighbors of a cell $c_1$ may not include $c_2$, 
which bears a certain resemblance to the preferential attachment models \cite{Barabasi}. As a result, the vertex degrees 
$k$ of the (undirected) relational graph may differ from one another and from $n_0$. A direct computational verification 
shows that $k$ is distributed according to a power law, $P(k) \sim k^{-\gamma}$, where $\gamma$ ranges, for different 
$n_0$, between $\gamma \sim 2$ and $\gamma \sim 4$  (Fig.~\ref{Statistics}), which implies that $G(n_0)$ demonstrates 
scale-free properties \cite{Barabasi,Albert} characteristic of the hippocampal network \cite{Li,Bonifazi}. In contrast, the 
histogram of the vertex degrees in the threshold-controlled relational graph $G(\theta)$ may be fit with the negative binomial 
distribution (Fig.~\ref{Statistics}B), which indicates that $G(\theta)$ is similar to a random graph. 

This neighbor-selection method for building the relational graph $G(n_0)$ has a number of other immediate advantages over 
the threshold-controlled construction of $G(\theta)$. For example, no cells are excluded from $\mathcal{T}_{0}$ due to the 
low appearance of the edges connecting to them. As a result, the simplex fields are distributed more uniformly 
(Fig.~\ref{SupplFigure6}), which helps capture the correct piecewise connectedness of the environment.

%%%%%%%%%%%%%%%%%%%%%%%%%%%%%%%%%%%
\begin{figure} 
\includegraphics[scale=0.8]{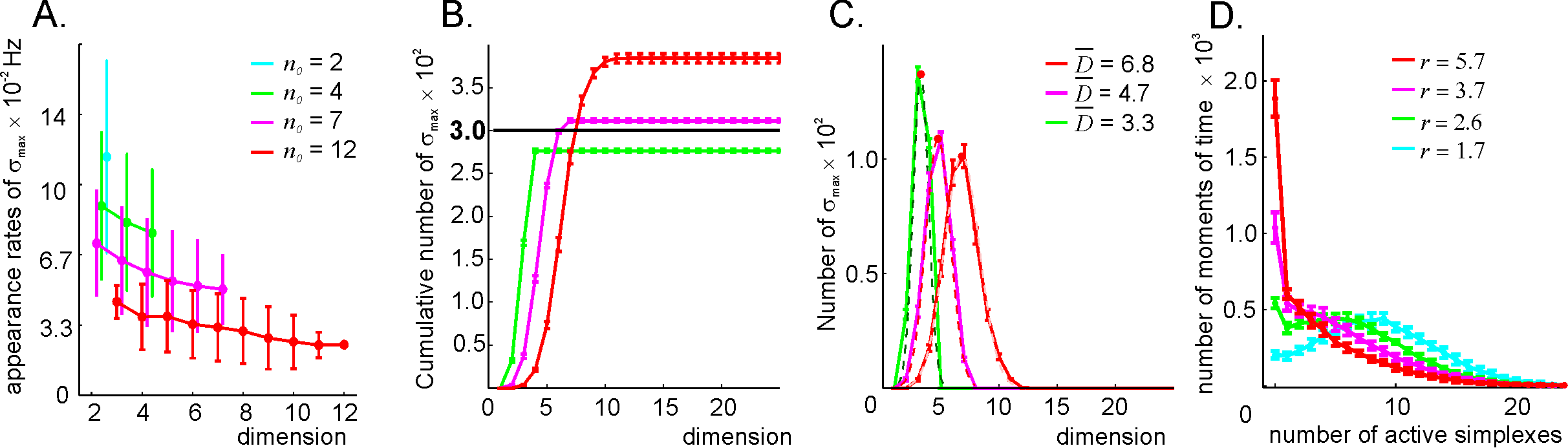}
\caption{\label{M2} \textbf{Figure 9. Selecting maximal simplexes via best neighbor selection (Method II)}. ({\bf A}) The 
appearance rates of the maximal simplexes in the simplicial complex $\mathcal{T}(n_0)$, computed for four different values 
of $n_0$  (color coded), decrease as a function of their dimensionality. ({\bf B}). Cumulative distribution of the number of 
maximal simplexes $N_{\max}$ over the selected simplexes' dimension. For the tested values of $n_0$, the fixed number 
of vertexes $N_c = 300$, indicated by the horizontal black line, is close to the number of maximal simplexes. For $n_0 = 7$, 
the values $N_{\max}$ and $N_c$ come closest. ({\bf C}). The histograms of the maximal simplexes' dimensionalities, fit to 
the normal distribution, indicate that for the relational graph with a similar number of links, the mean dimensionalities of the 
maximal simplexes are smaller than in in the complex built via the threshold-selection method. The width of the distributions 
is about $40\% $ of $\bar D$. ({\bf D}) The histogram of the number of coactive maximal simplexes, fit to an exponential 
distribution. An expected number of coactive simplexes ranges between $\beta = 2$ and $\beta = 6$. The procedure of 
averaging over the place field maps is the same.} 
\end{figure} 
%%%%%%%%%%%%%%%%%%%%%%%%%%%%%%%%%%%

By studying the properties of the clique complexes produced by the relational graphs $G(n_0)$ for $n_0 = 2$, 4, 7 and 
12---parameters chosen to produce similar numbers of edges as in the previous method---we  found that the number of 
maximal cliques in $\mathcal{T}_{0}(G(n_0))$ is typically lower than in $\mathcal{T}_{0}(G(\theta))$. The appearance 
rates of maximal cliques in $G(n_0)$ are more scattered and less sensitive to dimensionality than in $G(\theta)$ 
(Fig.~\ref{Statistics}A and Fig.~\ref{SupplFigure6}). The number of maximal simplexes in $\mathcal{T}_{0}(n_0)$ 
remains close to the number of  cells (Fig.~\ref{Statistics}B) and their dimensionality is lower than in the threshold-based 
selection approach (Fig.~\ref{Statistics}C). The contiguity index in all complexes ranges between to $\xi = 0.67$ and 
$\xi = 0.71$. The coverage of the space with the simplex fields improves with growing $n_0$ (see Fig.~\ref{SupplFigure6})---for 
$n_0 > 2$ the complex $\mathcal{T}_{0}(n_0)$ is connected, while the behavior of $b_0$ is more 
regular (see Table~\ref{SupplTable3}). However, the path connectivity of the complex $\mathcal{T}_{0}(n_0)$ remains deficient 
for all $n_0$ because the number of stable spurious $1D$ loops remains high (Table~\ref{SupplTable3}). After filling the gaps and 
closing the holes, most complexes constructed for $n_0 \geq 7$ acquire correct topological signatures (Table~\ref{SupplTable3}), 
and the requirements to $\mathcal{T}_{0}$ are satisfied. Thus, the simplicial complex obtained by the neighbor selection 
method for $n_0 \geq 7$ can also be viewed as a ``cell assembly complex,'' meaning it can serve as a formal model of the 
place cell assembly network with mean contiguity $\xi = 0.7$. 

\section{Discussion}
\label{section:discussion}

The proposed approach allows creating schematic models of the place cell assembly network---the cell assembly complex 
$\mathcal{T}_{0}$---by controlling the basic phenomenological parameters of place cell coactivity and then relating the 
network's architecture to the net topological information it encodes. The coactivity complex $\mathcal{T}$ was previously 
used for representing the pool of place cell coactivities \cite{Dabaghian,Arai,Curto}. Specifically, the low order (pair and 
triple) coactivity events were used to construct the $2D$ skeleton of $\mathcal{T}$, and then its $0D$ and $1D$ topological 
loops were matched with the corresponding topological signatures of the environment. In the current study, the entire pool 
of place cell coactivities is used to model the full place cell assembly network, including the higher order assemblies 
representing both the low-dimensional spatial environment as well as high-dimensional memory space \cite{ Buzsaki,Eichenbaum}. 
The learning times $T_{\min}$ estimated from the dynamics of the $0D$ and $1D$ loops in $\mathcal{T}_{0}$ remain close 
to the learning times computed for the full coactivity complex $\mathcal{T}$ (see Table~\ref{SupplTable4}). This implies that 
the selected, ``core'' pool of coactive place cell combinations captures the topological structure of the environment as fast 
and as reliably as the entire set of the place cell coactivities.

We view the proposed algorithms as basic models of a more general ``phenomenological'' approach, one which can be 
further developed along several broad lines. First, the structure of the relational graph is currently deduced from the activity 
vectors defined over the entire navigation period $T = 25$ minutes. A biologically more plausible selection algorithm should 
be adaptive: the structure of the relational graph at a given moment of time $t < T$ should be based only on the spiking 
information produced before $t$. Hence, in a more advanced model, the structure of the relational graph should develop 
in time, and in general the cell assemblies comprising $\mathcal{T}_{0}$ should be derived using synaptic and structural 
plasticity mechanisms. Second, the selection criteria in Methods I and II above may be individualized: the appearance threshold 
used to construct the relational graph can be assembly-specific, i.e. $\theta = \theta(\sigma)$, so that the properties of the 
resulting network would be described in terms of the probability distribution of the threshold values across the cell assembly 
population. Similarly, the number of closest neighbors can be made cell-specific, $n_0 = n_0(c_i)$, which should permit 
better control over the topological properties both of the network and of the cell assembly complex. Third, threshold control 
can be implemented using different coactivity metrics, for instance via the pairwise correlation coefficient
\begin{equation}
\rho(c_1,c_2)=(m_1 \cdot m_2)/|m_1||m_2|,
\label{metric}
\end{equation}
which would connect cells with correlated spiking (irrespective of their firing rates), in contrast with the metric (\ref{metric}), 
which does the opposite. In general, two metrics $\rho$ and $\rho'$, produce relational graphs with different topologies. 
Nevertheless, they may produce similar or identical large-scale effects, such as generating topologically identical cell 
assembly complexes $\mathcal{T}_{0}$, or exhibit similar learning times, $T_{\min}$. Identifying classes of metrics 
that produce topologically similar results will be examined in future research.

\section{Acknowledgments}
\label{section:acknow}

The work was supported in part by Houston Bioinformatics Endowment Fund, the W. M. 
Keck Foundation grant for pioneering research and by the NSF 1422438 grant.

\newpage
\beginsupplement

\section{Supplementary Figures}
\label{section:SupplFigs}
%%%%%%%%%%%%%%%%%%%%%%%%%%%%%%%%%%
\begin{figure}[ht] 
\includegraphics[scale=0.77]{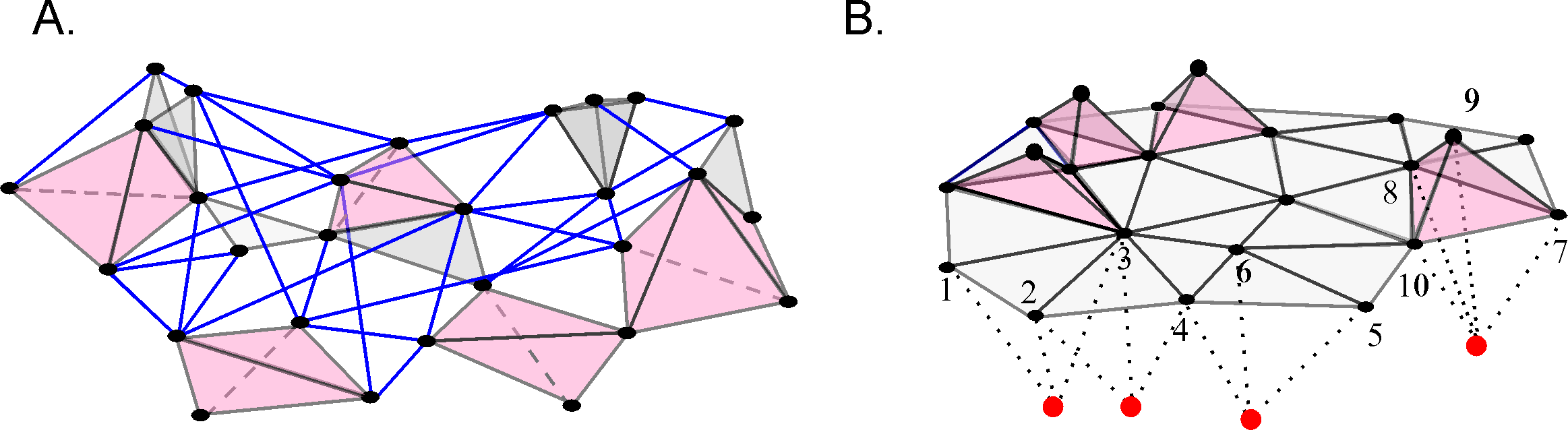}
\caption{\label{SupplFigure1} \textbf{Simplicial complexes}. ({\bf A}) A schematic representation of an irregular simplicial complex 
$\mathcal{K}$, in which the number of maximal simplexes is larger than the number of vertexes (black dots). The maximal $1D$ 
simplexes are shown as blue segments, the $2D$ simplexes as gray triangles and the $3D$ simplexes as pink tetrahedrons. 
({\bf B}) A simplicial ``quasi-manifold,'' $\mathcal{Q}$, which has a similar number of vertexes and maximal simplexes of different 
dimensionalities. If each maximal simplex, e.g. $(1,2,3)$ or $(10,7,8,9)$, corresponds to an assembly of place cells driving a read–out 
neuron (red dots), then $\mathcal{Q}$ is a cell assembly complex. Dotted lines represent synaptic connections from the place cells to 
the readout neuron.} 
\end{figure} 
%%%%%%%%%%%%%%%%%%%%%%%%%%%%%%%%%%%

\newpage

%%%%%%%%%%%%%%%%%%%%%%%%%%%%%%%%%%%
\begin{figure} 
\includegraphics[scale=0.77]{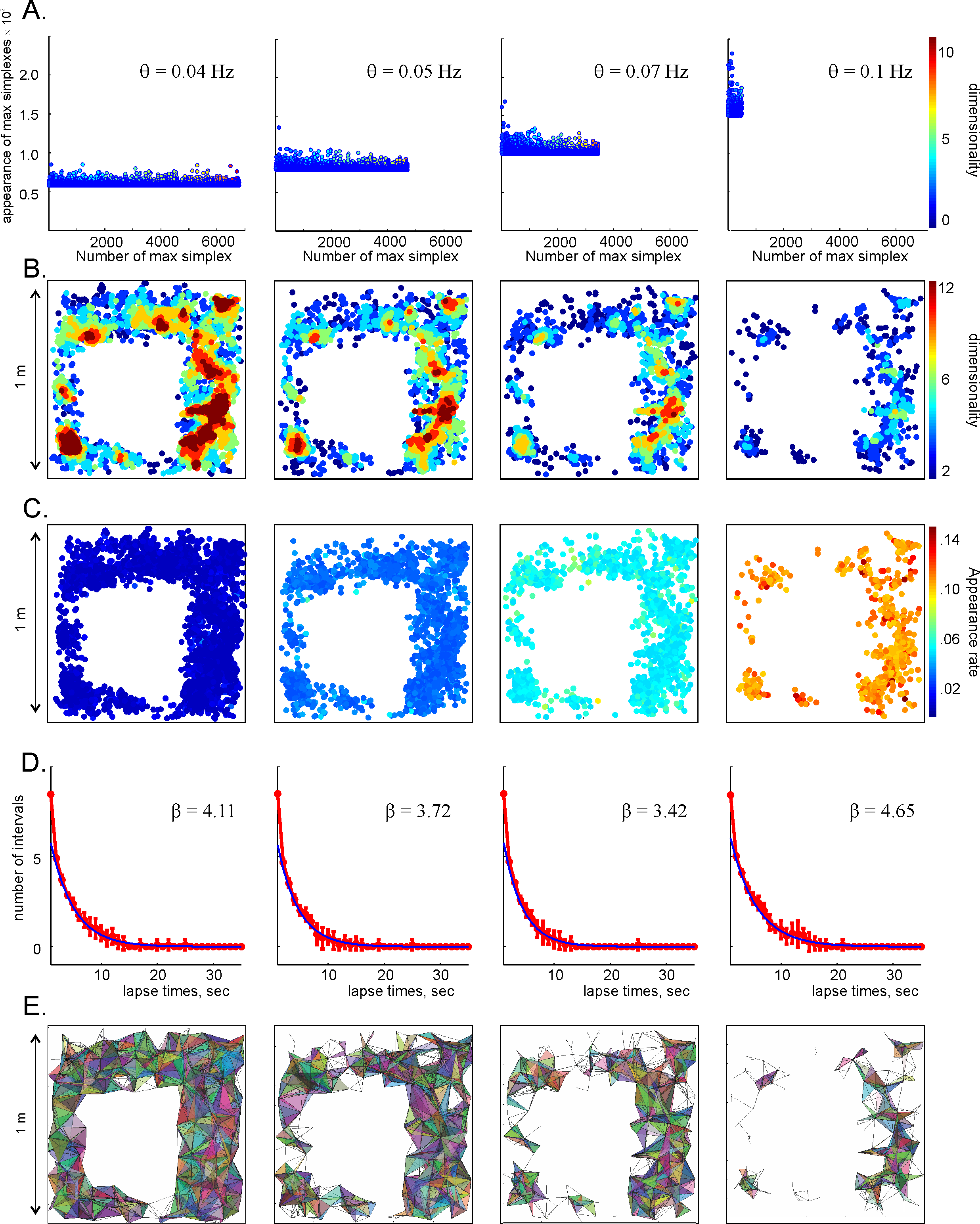}
\caption{\label{SupplFigure2} \textbf{The simplicial complexes $\mathcal{T}_0(\theta)$ constructed 
by direct selection of coactive combinations, for four different values of $\theta$}. ({\bf A}) The appearance rates of simplexes, 
arranged from left to right according to their dimension. Each dot corresponds to a maximal simplex whose dimension is 
color-coded according to the colorbar on the right. ({\bf B}) Spatial distribution of the dimensionalities of the selected simplexes. 
({\bf C}) Spatial distribution of the appearance rates of the selected simplexes. ({\bf D}) The histograms of the lapse times, fit 
to double exponential distribution (blue line), and the value of the fitted distribution's rate $\beta$. ({\bf E}) Spatial projections of 
the $2D$ skeletons of the $\mathcal{T}_{0}(\theta)$. Data for all panels is computed for a specific place field map for illustrative 
purposes.} 
\end{figure} 
%%%%%%%%%%%%%%%%%%%%%%%%%%%%%%%%%%%

\newpage

%%%%%%%%%%%%%%%%%%%%%%%%%%%%%%%%%%%
\begin{figure} 
\includegraphics[scale=0.77]{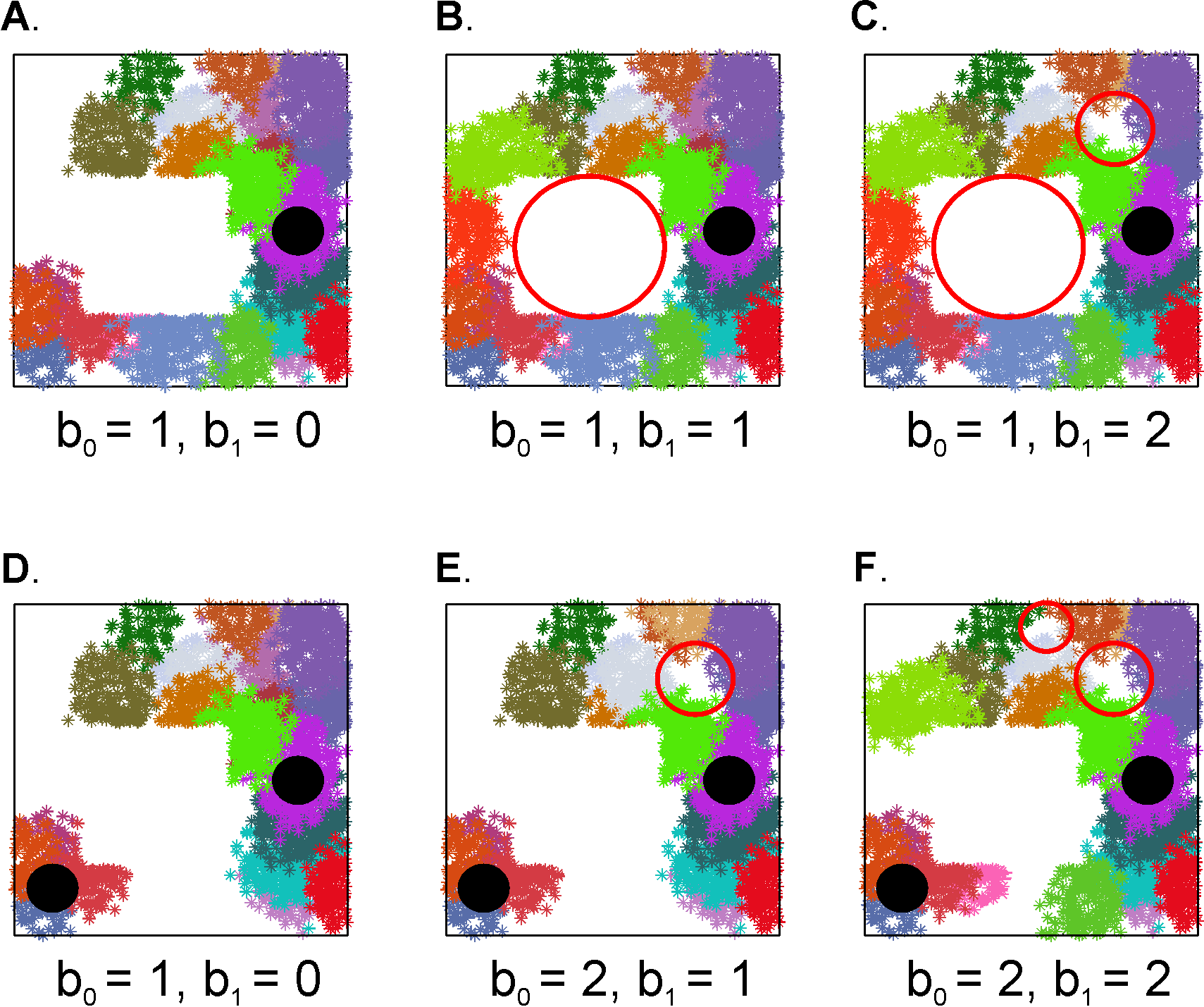}
\caption{\label{SupplFigure3} \textbf{Low-dimensional topological features of the cell assembly complexes captured by place 
field map}. ({\bf A}) A place filed map corresponding to a singly-connected ($b_0 = 1$, marked by the dot) complex with no 
non-contractible loops ($b_1 = 0$). ({\bf B}) A place field map corresponding to a complex with correct list of Betti numbers 
(correct topological barcode \cite{Ghrist}): the physical hole (see Figures 1 and 3) produces one 
non-contractible loop (red circle, $b_1 = 1$). ({\bf C}) A map containing a spurious hole in the upper-right corner produces 
an extra persistent loop (the additional small circle, net $b_1 = 2$). ({\bf D}) A map containing two disconnected pieces marked
by the black dots (net $b_0 = 2$) and having no non-contractible loops ($b_1 = 0$). ({\bf E}) A map containing two pieces and 
one persistent spurious $1D$ loop. ({\bf F}) The green and the brown place fields at the top connect, yielding another persistent 
spurious $1D$ loop. Compare these illustrations and the topological barcodes to the illustrations and Suppl. Movies provided in 
\cite{Singh}} 
\end{figure} 
%%%%%%%%%%%%%%%%%%%%%%%%%%%%%%%%%%%

\newpage

%%%%%%%%%%%%%%%%%%%%%%%%%%%%%%%%%%%
\begin{figure} 
\includegraphics[scale=0.77]{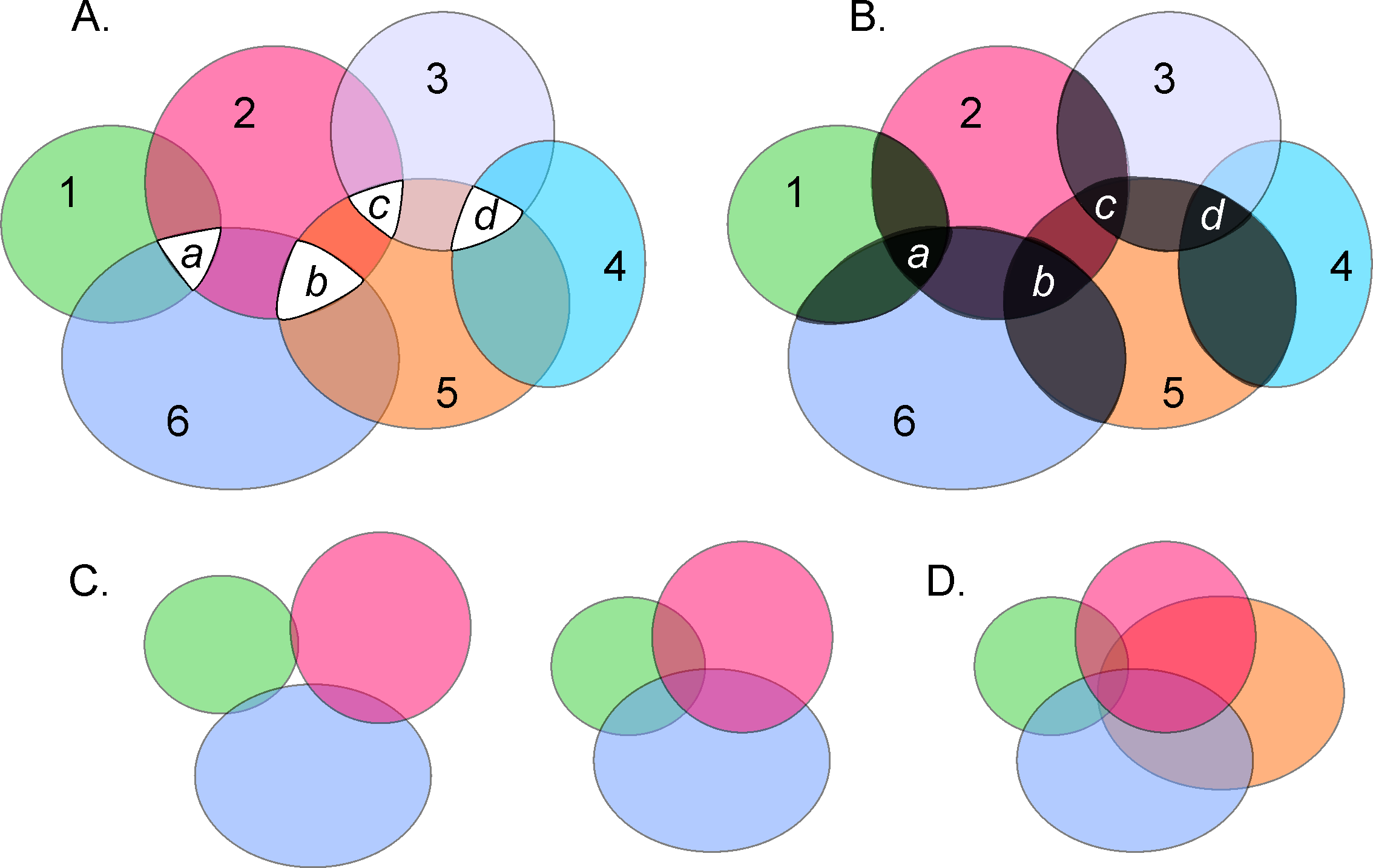}
\caption{\label{SupplFigure4} \textbf{Schematic illustrations of spatial maps}. ({\bf A}) Schematic representations of $3^d$ 
order simplex fields, $a$, $b$, $c$ and $d$, encoded by third order maximal simplexes, $\sigma_a = (v_1, v_2, v_6)$, 
$\sigma_c = (v_2, v_3, v_5)$, etc. ({\bf B}) If the $2D$ simplexes are discarded and their $1D$ faces are retained, then the 
second-order simplex fields are produced, shown here as overlapping shaded regions. The original simplex fields $a$, $b$, 
$c$ and $d$ are now represented by the coactivity of three pairs, e.g., a is represented by $\sigma_{a,1} = (v_1, v_2)$, 
$\sigma_{a,2} = (v_2, v_6)$ and $\sigma_{a,3} = (v_1, v_6)$. ({\bf C}). The three place fields on the left exhibit pairwise, 
but not triple overlap. In the generic spatial configuration shown on the right, pairwise overlapping place fields also produce a 
triple overlap. ({\bf D}) Four pairwise overlapping convex regions in $2D$ produce all the higher order (triple and quadruple) 
overlaps.} 
\end{figure} 
%%%%%%%%%%%%%%%%%%%%%%%%%%%%%%%%%%%

\newpage

%%%%%%%%%%%%%%%%%%%%%%%%%%%%%%%%%%%
\begin{figure} 
\includegraphics[scale=0.77]{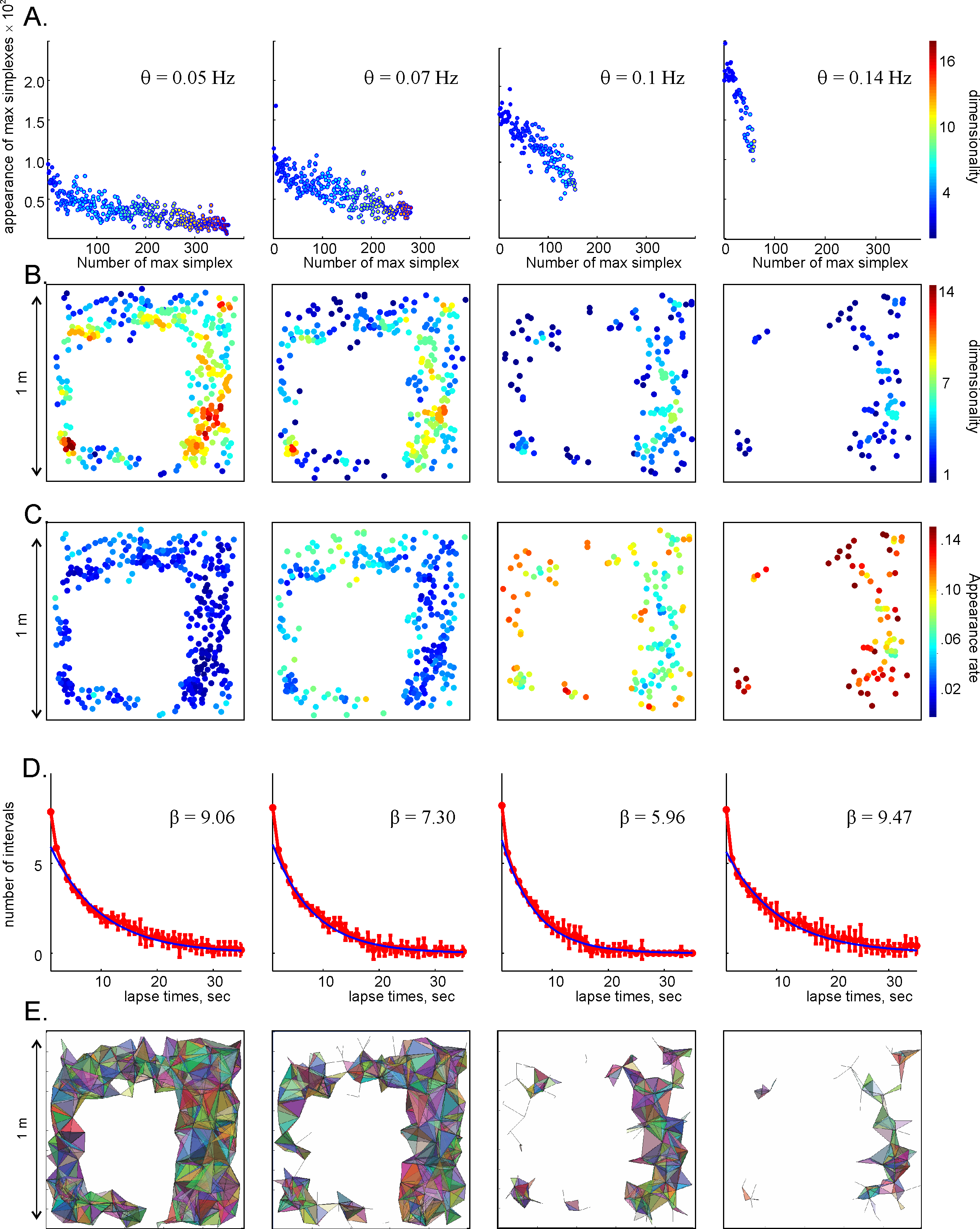}
\caption{\label{SupplFigure5} \textbf{Simplicial complexes $\mathcal{T}_0(\theta)$ constructed via 
pairwise coactivity selection (Method I) for four different threshold values}. ({\bf A}) The appearance rates of the maximal 
simplexes, arranged according to their dimension, demonstrate remarkably tight, graph-like distribution. The color of the dots 
corresponds to the dimension of the simplexes, as indicated by the colorbar on the right. ({\bf B}) Spatial distribution of the 
dimensionalities of the selected simplexes. ({\bf C}) Spatial distribution of the appearance rates of the selected simplexes. 
({\bf D}) The histograms of the lapse times, fit to double exponential distribution, and the value of the fitted distribution's 
parameter $\beta$. ({\bf E}) Spatial projections of the $2D$ skeletons of $\mathcal{T}_{0}(\theta)$. Data for all panels is 
computed for a specific place field map for illustrative purposes.} 
\end{figure} 
%%%%%%%%%%%%%%%%%%%%%%%%%%%%%%%%%%%

\newpage

%%%%%%%%%%%%%%%%%%%%%%%%%%%%%%%%%%%
\begin{figure} 
\includegraphics[scale=0.77]{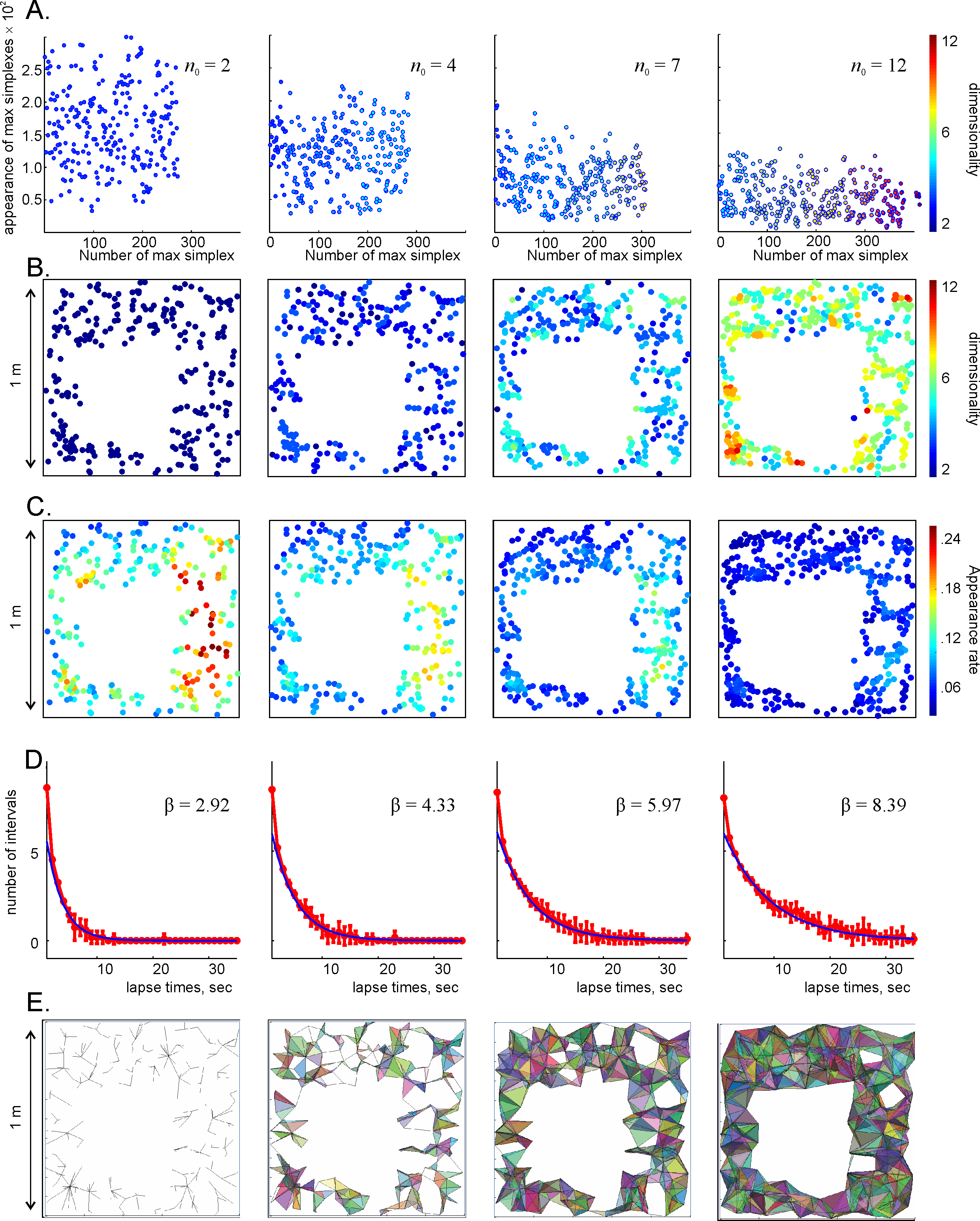}
\caption{\label{SupplFigure6} \textbf{The selected simplicial complexes $\mathcal{T}_0(n_0)$ 
constructed via the closed-neighbor selection algorithm (Method II) for four different values of $n_0$}. ({\bf A}) The 
appearance rates of simplexes, arranged according to their dimensions, color-coded as indicated by the colorbar on the 
right. ({\bf B}) Spatial distribution of the dimensionalities of the selected simplexes. ({\bf C}) Spatial distribution of the 
appearance rates of the selected simplexes. ({\bf D}) The histograms of the lapse times, fit to double exponential distribution, 
and the value of the fitted distribution's parameter $\beta$. ({\bf E}) Spatial projections of the $2D$ skeletons of $\mathcal{T}_{0}(n_0)$.} 
\end{figure} 
%%%%%%%%%%%%%%%%%%%%%%%%%%%%%%%%%%%%%%%%%%%%%%%%%%%%%%%%%%%%%%%%%%%%%%%%%%

%%%%%%%%%%%%%%%%%%%%%%%%%%%%%%%%%%%%%%%%%%%%%%%%%%%%%%%%%%%%%%%%%%%%%%%%%%
\pagebreak
\clearpage
\newpage
%\textbf{SUPPLEMENTARY TABLES}
\section{Supplementary Tables}
\label{section:tables}
\vspace{5mm}
%%%%%%%%%%%%%%%%%%%%%%%%%%%%%%%%%%%%%%%%%%%%%%%
\begin{table}[ht]
\begin{tabular}{| c | c | c | c |}
  \cline{1-4}
  \hline
%  \multicolumn{1}{}{} 
{\bf $\theta$ = 0.04 Hz} & {\bf $\theta$ = 0.05 Hz} & {\bf $\theta$ = 0.07 Hz} & {\bf $\theta$ = 0.1 Hz}  \\  \hline
1 139 509 569 	&1 171 393 410 	&2 198 324 214 	&5 131 108 15 \\
1 126 409 480 	&1 175 353 286 	&1 203 247 144 	&7 147 71  17 \\
1 156 580 853 	&2 211 518 571 	&2 233 330 419 	&6 139 152 41 \\
1 135 407 551 	&1 176 369 333 	&1 201 296 152 	&7 139 50 7 \\
1 149 422 585 	&2 172 342 453 	&2 185 312 320 	&13 122 107 73\\ 
1 130 485 775 	&1 186 447 392 	&1 190 335 250 	&4 164 84 27 \\
1 106 459 655 	&1 202 352 603 	&1 177 395 353 	&8 179 117 25 \\
1 120 519 787 	&1 194 397 449 	&1 215 333 237 	&8 160 109 63 \\
1 131 441 757 	&1 165 459 499 	&1 203 339 197 	&8 149 65 15 \\
1 160 364 575 	&1 226 313 368 	&1 175 270 222 	&8 117 92 14 \\ \hline
\end{tabular}
\caption{\textbf{Topological signature of the complex selected by simplex rate thresholding}. Each 
cell contains a list of four Betti numbers, ($b_0$, $b_1$, $b_2$, $b_3$). For the low rate, $f_{\sigma} = 0.04$ Hz, 
$\mathcal{T}_{0}(\theta)$ encodes the correct spatial connectedness of the environment ($b_0 = 1$). For 
intermediate rates, $0.05 \leq f_{\sigma} \leq 0.07$ Hz, $\mathcal{T}_{0}(\theta)$ may occasionally break into 
two pieces ($b_0 = 2$) and for $f_{\sigma} \geq 0.10$ Hz and higher, $\mathcal{T}_{0}(\theta)$ fragments into 
multiple components. In higher dimensions $D \geq 1$, $\mathcal{T}_{0}(\theta)$ contains over a hundred 
noncontractible topological loops.}\label{SupplTable1}
\end{table}
%%%%%%%%%%%%%%%%%%%%%%%%%%%%%%%%%%%%%%%%%%%%%%%

%%%%%%%%%%%%%%%%%%%%%%%%%%%%%%%%%%%%%%%%%%%%%%%
\begin{table}[ht]
 A. Original  \hspace{21mm}	B. Corrected   \\ %\hline
 \vspace{0.5mm}
\begin{tabular}{| c | c | c | c | c | c | c | c | c |}
%\cline{1}\cline{4}\cline{6}\cline{9}
% \multicolumn{1-4}{|c|}{A. Original}\multicolumn{5-9}{|c|}{A. Original} \\
 %& & A. Original  & & & &	B. Corrected  & &  \\ %\hline
\cline{1-4}\cline{6-9}
{\bf $\theta$ = 0.05} & {\bf $\theta$ = 0.07} & {\bf $\theta$ = 0.1} & {\bf $\theta$ = 0.14} &  & {\bf $\theta$ = 0.05} & {\bf $\theta$ = 0.07} & {\bf $\theta$ = 0.1} & {\bf $\theta$ = 0.14} \\  %\hline
\cline{1-4}\cline{6-9}
\textbf{1 1 0 0} 	&2 1 0 0 	&5 1 0 0 	&3 0 0 0  & 	&\textbf{1 1 0 0} 	&\textbf{1 1 0 0} 	&1 0 0 0 	&3 0 0 0 \\
1 2 0 0 	&1 5 0 0  	&7 2 0 0  	&7 1 0 0  & 	&\textbf{1 1 0 0} 	&\textbf{1 1 0 0} 	&1 0 0 0 	&2 0 0 0 \\
1 2 0 0 	&2 1 0 0 	&6 2 0 0  	&7 2 0 0  & 	&\textbf{1 1 0 0} 	&\textbf{1 1 0 0} 	&1 0 0 0 	&1 0 0 0 \\
\textbf{1 1 0 0} 	&1 2 0 0  	&7 7 0 0  	&7 1 0 0  & 	&\textbf{1 1 0 0} 	&\textbf{1 1 0 0} 	&1 0 0 0 	&1 0 0 0 \\
\textbf{1 1 0 0} 	&2 3 0 0 	&13 1 0 0 	&8 1 0 0  & 	&\textbf{1 1 0 0} 	&\textbf{1 1 0 0} 	&1 0 0 0 	&3 0 0 0 \\
1 3 0 0 	&1 0 0 0 	&4 6 0 0 	&11 0 0 0  & 	&\textbf{1 1 0 0} 	&\textbf{1 1 0 0} 	&1 0 0 0 	&1 0 0 0 \\
2 1 0 0 	&1 5 0 0 	&8 1 0 0  	&7 5 0 0  & 	&\textbf{1 1 0 0} 	&\textbf{1 1 0 0} 	&2 0 0 0 	&3 0 0 0 \\
1 2 0 0 	&1 2 0 0 	&8 2 0 0  	&9 1 0 0  & 	&\textbf{1 1 0 0} 	&\textbf{1 1 0 0} 	&1 0 0 1 	&2 0 0 0 \\
1 3 0 0 	&1 4 0 0 	&8 2 0 0  	&9 2 0 0  & 	&\textbf{1 1 0 0} 	&1 1 1 0 	&\textbf{1 1 0 0} 	&4 0 0 0 \\
1 3 0 0 	&1 5 0 0 	&8 2 0 0  	&7 6 0 0  & 	&\textbf{1 1 0 0} 	&\textbf{1 1 0 0} 	&2 0 0 0 	&3 0 0 0  \\ \hline
\end{tabular}
\caption{\textbf{Topological signature of the complex selected by link-rate thresholding}. 
({\bf A}) For the low rate $f_{\sigma} = 0.05$ Hz, $\mathcal{T}_{0}(\theta)$ occasionally produces the 
correct topological signature ($b_0 = b_1 = 1$, $b_{n > 1} = 0$, shown in bold). For intermediate rates 
$f_{\sigma} \sim 0.07$ Hz, $\mathcal{T}_{0}(\theta)$ may occasionally break into two pieces ($b_0 = 2$) 
and produce extra noncontractible loops in $1D$. For high thresholds, $f_{\sigma}\geq 0.10$ Hz, 
$\mathcal{T}_{0}(\theta)$ fragments into multiple components. However, the connectivity of 
$\mathcal{T}_{0}(\theta)$ in higher dimensions, $D \geq 2$, is correct for all cases, which implies that 
$\mathcal{T}_{0}(\theta)$ contracts into $2D$. ({\bf B}) After applying the correction algorithms, the 
selected complexes acquire correct topological signature for $f_{\sigma}\leq 0.07$ for all maps. The 
corresponding learning times $T_{\min}$ are listed in Suppl. Table 4.}\label{SupplTable2}
\end{table}
%%%%%%%%%%%%%%%%%%%%%%%%%%%%%%%%%%%%%%%%%%%%%%%

\begin{table}[ht]
 A. Original  \hspace{21mm}	B. Corrected   \\ %\hline
 \vspace{0.5mm}
\begin{tabular}{| c | c | c | c | c | c | c | c | c |}
%\cline{1}\cline{4}\cline{6}\cline{9}
% \multicolumn{1-4}{|c|}{A. Original}\multicolumn{5-9}{|c|}{A. Original} \\
\cline{1-4}\cline{6-9}
%A.	Original	B.	Corrected
{\bf $n_0$ = 2} & {\bf $n_0$ = 5} & {\bf $n_0$ = 7} & {\bf $n_0$ = 12} &  & {\bf $n_0$ = 2} & {\bf $n_0$ = 5} & {\bf $n_0$ = 7} & {\bf $n_0$ = 12} \\ \hline
29 0 	&1 31 0 0 	&1 11 0 0 	&1 4 1 0 	&  &1 2 	&\textbf{1 1 3 0} 	&\textbf{1 1 1 0} 	&\textbf{1 1 1 1} \\
29 0 	&1 30 0 0 	&1 16 0 0 	&1 3 0 0 	&  &\textbf{1 1} 	&1 2 0 0 	&\textbf{1 1 1 0} 	&\textbf{1 1 1 0} \\
32 0 	&1 28 0 0 	&1 19 0 0 	&1 3 0 0 	&  &1 4 	&1 2 0 0 	&\textbf{1 1 0 1} 	&\textbf{1 1 1 0} \\
31 0 	&1 31 0 0 	&1 14 0 0 	&1 2 0 0 	&  &1 3 	&1 4 0 0 	&\textbf{1 1 1 0} 	&\textbf{1 1 1 0} \\
31 0 	&2 34 0 0 	&1 20 0 0 	&1 4 0 0 	&  &1 3 	&1 2 0 0 	&\textbf{1 1 1 0} 	&1 2 2 1 \\
34 0 	&1 26 0 0 	&1 12 0 0 	&1 2 0 0 	&  &1 2 	&1 5 0 0 	&\textbf{1 1 1 0} 	&1 2 0 1 \\ 
28 0 	&1 41 0 0 	&1 11 0 0 	&1 2 0 0 	&  &1 4 	&1 2 0 0 	&1 2 1 0 	&1 2 0 0 \\ 
28 0 	&1 40 0 0 	&1 15 0 0 	&\textbf{1 1 1 0} 	&  &1 3 	&\textbf{1 1 0 0} 	&\textbf{1 1 2 0} 	&\textbf{1 1 0 2} \\
31 0 	&1 32 0 0 	&1 17 0 0 	&1 4 0 0 	&  &1 3 	&1 2 0 0 	&1 3 1 0 	&\textbf{1 1 0 0} \\
43 0 	&1 33 0 0 	&1 16 4 0 	&1 4 0 0 	&  &1 3 	&1 2 0 0 	&\textbf{1 1 5 0} 	&1 2 0 0 \\ \hline
\end{tabular}
\caption{\textbf{Topological signatures of the complexes selected by the neighbor-selection algorithm}. 
({\bf A}) If but one pair of closest vertexes is selected $n_0 = 2$, $\mathcal{T}_{0}(n_0)$ breaks into multiple 
components. For $n_0 \geq 5$, $\mathcal{T}_{0}(n_0)$ has only one component, but path connectivity is compromised 
($b_0 =1$, $b_1 \gg 1$). In higher dimensions, $\mathcal{T}_{0}(n_0)$ is contractible, $b_{n > 1} = 0$. 
({\bf B}) After applying the correction algorithms, the selected complexes for almost all maps acquire correct topological signature 
in $1D$ and $2D$ (shown in boldface) for $n_0 \geq 7$. The corresponding learning times, $T_{\min}$, are listed in Suppl. Table 4.}
\label{SupplTable3}
\end{table}

\begin{table}[ht]
\begin{tabular}{| c | c | c | c | c | c | c | c | c |c | c | c |}
\hline
	&1	&2	&3	&4	&5	&6	&7	&8	&9	&10	&mean/std \\ \hline
$\mathcal{T}_{0}$ 	&4.4 	&2.7 	&2.3 	&2.7 	&2.8 	&2.7 	&2.1 	&3.8	&10.7 	&2.7	 &3.7/2.5 \\
$\mathcal{T}_{0}(\theta =.05)$ 	&3.8 	&2.7	&1.9 	&2.7 	&3.6 	&3.8 	&3.8 	&3.8 	&2.5 	&3.7	 &3.2/0.7 \\
$\mathcal{T}_{0}(\theta =.07)$ 	&2.1 	&3.8 	&3.8 	&2.8 	&2.8 	&2.3 	&2.4 	&3.7 	&2.5 	&3.8	 &3.0/0.7 \\
$\mathcal{T}_{0}(n_0=7)$	&1.9 	&4.6 	&2.4 	&2.7 	&2.8 	&3.8	&$\infty$ 	&3.8	&$\infty$ 	&3.8	 &3.2/0.9 \\
$\mathcal{T}_{0}(n_0=12)$ 	&3.8 	&2.7	&1.9 	&2.7	&$\infty$	&$\infty$	&$\infty$ 	&3.8 	&3.8	&$\infty$	& 3.1/0.8 \\ 
\hline
\end{tabular}
\caption{\textbf{The learning times}, $T_{\min}$ (in minutes) computed for the selected simplicial 
complexes $\mathcal{T}_{0}(\theta)$ and $\mathcal{T}_{0}(n_0)$ are similar to the learning times computed via 
the full temporal nerve complex $\mathcal{T}$. Thus, the information about the topological structure of the environment 
emerges from the cell assembly activity as fast as from the entire pool of place cell coactivities. However, note that 
the complex $\mathcal{T}_{0}(n_0)$ sometimes fails to produce a finite leaning time; non-convergent cases are 
marked by $\infty$.}
\label{SupplTable4}
\end{table}

\newpage

%\textbf{SUPPLEMENTARY MOVIES ILLUSTRATING THE FIRST OF THE TEN TESTED MAPS}
\section{SUPPLEMENTARY MOVIES ILLUSTRATING THE FIRST OF THE TEN TESTED MAPS}
\label{section:movies}

\textbf{Suppl. Movie 1}. The grey dots represent centers of the place fields, viewed from above, similar to Figure 3B. The centers of the place fields that correspond to the coactive place cells are shown in red. The resulting “activity packet” moves in the environment following the simulated rat's trajectory. 

\textbf{Suppl. Movie 2}. Selection of the cell assemblies by Method I ($\theta=100$) and assigning readout neurons to the cell assemblies.

\textbf{Suppl. Movie 3}. A side projection view of the activity packet propagating in the cell assembly network, selected via Method I. To emphasize that the coactive place cell combinations comprise a cell assembly complex $\mathcal{T}_{0}(\theta)$, the corresponding place field centers are schematically connected to the readout neurons.

\textbf{Suppl. Movie 4}. The same system shown in the same projection as the Figure 3B and Suppl. Movie 1.

\end{document}